\documentclass[11pt,a4paper]{article}

\usepackage[english]{babel}
\usepackage[utf8]{inputenc}
\usepackage[left=2.5cm,right=2.5cm,top=3cm,bottom=3cm]{geometry}
\usepackage{amsfonts}
\usepackage{amsmath}
\usepackage{booktabs}
\usepackage{caption}
\usepackage{amssymb}
\usepackage{mathrsfs}
\usepackage{amsfonts}
\usepackage{mathtools}
\usepackage{tensor}
\usepackage{graphicx}
\usepackage{appendix}
\usepackage{color}
\usepackage[dvipsnames]{xcolor}
\usepackage{longtable}
\usepackage{array}
\usepackage{booktabs} \usepackage{ragged2e}

\usepackage[affil-it]{authblk}
\usepackage[backend=bibtex,style=numeric-comp,maxnames=5,sorting=none,url=false,hyperref=true,eprint=false,doi=true]{biblatex}
\usepackage{hyperref}
\hypersetup{
	colorlinks=true,
	urlcolor=blue,
	linkcolor={black!40!black},
	citecolor={green!40!black}
}
\usepackage{cleveref}

\def \blue  {\rm \color{blue} }

\newcommand{\dd}{{\rm{d}}}

\newcommand{\defeq}{:=}
\newcommand{\defeqs}{\stackrel{\scri}{\vcentcolon=}}
\def\scri{{\!\mathscr{J}}}

\newcommand{\im}{\mathrm{i}}
\renewcommand{\P}{{\mathcal{P}}}

\newcommand{\prn}[1]{\left(#1\right)}

\newcommand{\brkt}[1]{\left[#1\right]}

\newcommand{\cbrkt}[1]{\left\lbrace#1\right\rbrace}
\newcommand{\bcbrkt}[1]{\Big\lbrace#1\Big\rbrace}

\newcommand{\abs}[1]{\left|#1\right|}
\newcommand{\ct}[2]{\tensor{{#1}}{#2}} 

\title{\bf
Study of gravitational radiation at $\scri$\\ for the general type-D Einstein-Maxwell solution\\ with cosmological constant}

\author{Francisco Fernández-Álvarez\thanks{francisco.fernandez@ehu.eus}\ }
\author{José M. M. Senovilla\thanks{josemm.senovilla@ehu.eus}}
\affil[]{Departamento de Física\protect\\ Universidad del País Vasco UPV/EHU\protect\\ Apartado 644, 48080 Bilbao, Spain\protect\\\ }

\date{\today{}}

\begin{document}

\maketitle

\begin{abstract}
We characterize the gravitational radiation present in the general Pleba\'nski-Demia\'nski type~D solution with cosmological constant $\Lambda$ of any sign. We show that radiation escapes towards infinity if and only if the acceleration parameter is non-zero. We calculate the Cotton-York tensor and the boundary stress-energy tensor at infinity for negative $\Lambda$, and for positive $\Lambda$ the $D$ tensor that complements initial or final data at $\scri$. In addition, we recast a recent parametrization of this family of metrics into a new form by a simple homothety, correctly identifying $\Lambda$, and study the existence and regularity of the possible axes of symmetry, searching for a relation between deficit angles and radiation. We also identify the branch of general non-radiating solutions with non-vanishing Cotton-York tensor at $ \scri $.
\end{abstract}

\newpage

\section{Introduction}
	The detection of gravitational waves \cite{LigoVirgoPRL} has pushed a great effort in developing the theory of gravitational radiation in a very broad sense.  This gravitational wave astronomy, together with the observational discoveries on black holes \cite{EHT:2019}, have brought confirmation of many aspects of the theory developed decades ago \cite{Bondi1962,Sachs1962,Pirani57,Bel1958,Penrose62}. Nevertheless, there are important open questions, like the inclusion of a cosmological constant $\Lambda$ in the theory \cite{Penrose2011}, which has motivated a large set of proposals with different approaches \cite{Ashtekar2014,Szabados2015,Compere2019,Poole2019}.  In parallel, the development of exact solutions has been very fruitful and offers many models of generalised black hole solutions. The accelerating solutions \cite{Plebanski-Demianski:1976,ChngMannStelea2006,PodolskyVratny2020} are of special interest for the study of radiation \cite{KrtousPodolsky:2003tc,PodolskyOrtaggioKrtous:2003,Podolsky-Kadlecova:2009}. As an example, the C-metric ---see, e.g., \cite{Griffiths-Podolsky2009}--- came to be very useful in testing the ideas of gravitational radiation in full General Relativity for   asymptotically flat spacetimes \cite{Ashtekar-Dray81}. \\
	
	In a series of recent works \cite{Fernandez-Alvarez-Senovilla2022a, Fernandez-Senovilla:2022b,FernandezSenovilla:2026} a new formalism was developed to study the asymptotic structure of spacetimes in connection to the presence of gravitational radiation at infinity in full General Relativity and for any value of $\Lambda$ ---see  also references therein and \cite{Fernandez-Senovilla:2025,Senovilla2022} for  reviews. The ideas were tested with several exact solutions for any $\Lambda\geq 0$, and also in a separate work \cite{FernandezPodolskySenovilla:2024} for type-D metrics with $\Lambda>0$. For $ \Lambda<0 $, they have been exploited in holography  \cite{CiambelliPasterskiTabor:2024,ArenasCiambelliDiazJiaRivera:2026} and fluid/gravity correspondence with non-perfect Carrollian fluids, testing the framework on Robinson-Trautman solutions \cite{Diaz:2026} ---for previous studies on this kind of space-times in the context of holography and Carroll fluids see, e.g., \cite{Gath2015,Ciambelli2017,Ciambelli2018}. 
	In  \cite{FernandezPodolskySenovilla:2024}, a version of the general accelerating type-D black holes of the Pleba\'nski-Demia\'nski form largely studied in \cite{PodolskyVratny2021,PodolskyVratny2023} was used. A new version of the metric was put forward in \cite{Astorino:2024b} which included the non--rotating accelerating solution, as well as the remaining subcases. Shortly afterwards, the authors of \cite{OvcharenkoPodolskyAstorino:2024,OvcharenkoPodolskyAstorino2025} presented a compact form of the metric that allows for an easier study of its properties, as well as of its asymptotics, and will be further developed elsewhere \cite{Podolsky-toappear}.
	
	In the present work we use this new family of metrics to illustrate our ideas on gravitational radiation for any $\Lambda$. This includes the computation of the  asymptotic super-Poynting vector field, and the geometric characterization in terms of the principal null directions. We conclude that these spacetimes contain gravitational radiation at infinity if and only if the parameter interpreted as acceleration is non-vanishing. In addition, the Cotton-York tensor and the holographic and initial data TT-tensor for several subcases are computed explicitly for $\Lambda<0$ and $\Lambda>0$, respectively. For $\Lambda >0$ this provides the spacetime initial or final data at $\scri$. We study other properties of the metric as well. In particular, we present a systematic study of the existence of axes of symmetry and of the possible conical singularities, in general, and analyze the relevant subcases. This allows us to connect the existence of gravitational radiation with the possibility of regularizing the deficit angle, or not at two components of the axis simultaneously. 

\section{The metric with $\Lambda$}
\label{sectionc:new-metric}

	In \cite{Astorino:2024b}, Astorino presented a new metric representation of the most general type~D solution of the Einstein-Maxwell field equations with any cosmological constant $\Lambda$ and a doubly-aligned, non-null electromagnetic field. Subsequently, in \cite{OvcharenkoPodolskyAstorino:2024} such metric was recast into a more compact form, denoted as~A$^+$ where a constant named $C_f$ was treated as a free gauge. However one can prove that this constant is a true parameter of the solution that can be adjusted according to the regularity of possible axes of symmetry ---see subsection \ref{subsec:cones}. This introduced a subtle shift in the identification of the cosmological constant. Herein, we present the corrected version of the A$^+$ solution and, by means of a simple homothety, resolve any ambiguity regarding the interpretation of the constant $C_f$ and the proper $\Lambda$ : \footnote{See \cref{sec:metric-correction}.}
	
\begin{equation}
    \dd \hat{s}^2=\dfrac{1}{\Omega^2}\bigg[-\dfrac{Q}{C_f\rho^2}(A\,\dd t - B\,\dd\varphi)^2
        + \dfrac{P}{C_f\rho^2}(C\,\dd t + D\,\dd\varphi)^2
        + \,\rho^2 \Big(\,\dfrac{\dd q^2}{Q} + \dfrac{\dd x^2}{P}\,\Big)\bigg]
    \label{A-new-metric}
\end{equation}
with 
\begin{align}
    \Omega &= q-\alpha\, x \,,\label{Omega_new}\\
    A &= 1 + \alpha^2(l^2-a^2)\,x^2\,,\label{eq:A}\\
    B &= a + 2l\,x + a\,x^2,\label{eq:B}\\
    C &= \alpha^2 a + 2\alpha l\,q + a\, q^2\,,\label{eq:C}\\
    D &= 1 + (l^2-a^2)\,q^2\,,\label{eq:D} \\
    \rho^2 &= AD+BC \,, \label{rho2_new}
\end{align}
and
	\begin{align}
	Q(q) &= (q^2-\alpha^2)\brkt{1-2mq + (a^2+e^2+g^2-l^2)q^2} \nonumber\\
	   &  - \frac{\Lambda}{3}\brkt{\,1+\alpha^2a^2 + 4\alpha a lq + 3l^2q^2
	     } ,\label{eq:Q-hom} \\
	P(x) &= (1-x^2)\big[1-2\alpha m x + \alpha^2(a^2+e^2+g^2-l^2) x^2\big] \nonumber\\
	   & - \frac{\Lambda}{3}\cbrkt{3l^2 x^2
	    +4al x^3 + \brkt{a^2+\alpha^2(a^2-l^2)^2} x^4 } . \label{eq:P-hom}
	\end{align}

The metric \cref{A-new-metric} is a solution of EFE with electromagnetic vector potential $A=A_t \dd t + A_\varphi \dd \varphi$ that reads\footnote{We have adapted it from \cite{OvcharenkoPodolskyAstorino:2024} to the coordinate $q$ and set the correct gauge ---see \cref{sec:metric-correction}.}

\begin{align}
A_t=&-\frac{1}{\sqrt{C_f}l\rho^2}\sqrt{\frac{1+\alpha^2\prn{a^2-l^2}}{1+\alpha^2a^2}}\bcbrkt{\prn{1+\alpha^2a^2+\alpha alq}g + elq+l\prn{g-\alpha a e}\prn{a\alpha^2+2\alpha l q+ a q^2}x\nonumber\\
+&\prn{\alpha l+aq}\brkt{\prn{ag-el^2\alpha}q+\alpha l g+ \alpha^2 a \brkt{\prn{a^2-l^2}gq-el}}x^2} + \frac{g}{\sqrt{C_f}l}\ ,\label{eq:vector-potential-t}\\
A_\varphi=&\frac{x}{\sqrt{C_f}\rho^2}\sqrt{\frac{1+\alpha^2\prn{a^2-l^2}}{1+\alpha^2a^2}}\bcbrkt{\alpha l^3\prn{e+\alpha a g}xq^2+l^2\brkt{\alpha^2a\prn{e+\alpha a g}xq+\prn{\alpha a l-g}q^2}\nonumber\\
+&\prn{1+a^2\alpha^2}\brkt{a e\prn{\alpha+xq}-galxq^2+g\prn{1-\alpha a^2 xq}+el\prn{2q+\alpha x}}}-aA_t\label{eq:vector-potential-phi}
\end{align}
This metric possesses two commuting Killing vectors $\partial_t$ and $\partial_\varphi$, the latter is generically cyclic --however, see footnote \ref{foot}.
To keep the $(-,+,+,+)$ signature of the metric the condition $P(x)>0$ is indispensable. Then, the coordinate system is defined only at either $Q(q)>0$ or $Q(q)<0$, the latter with $q$ as a time coordinate and the former with $t$ as such. One can check that $Q(q)=0$ are just coordinate singularities that represent horizons through which the metric can be extended joining regions with both signs of $Q$. In the extended metric, these become Killing horizons for the Killing vector field (KVF)
	\begin{equation}
		\xi^\alpha=\omega \delta^\alpha_{t} +\tau\delta^\alpha_{\varphi}\ ,\label{eq:generalKVF}
	\end{equation}
with $\omega$ and $\tau$ constants that satisfy 
$$
\omega C(q_0) +\tau D(q_0) =0
$$
at the root $q_0$ of $Q$: $Q(q_0)=0$.

Hence, the ranges of the coordinates are:  $t\in (-\infty,\infty)$;\footnote{Unless both identifications in \eqref{eq:identify} are simultaneously performed.} $\varphi$ is intended to be a cyclic coordinate whose total length depends on the value of $C_f$, see subsection \ref{subsec:cones}. Nevertheless, when $P(x) >0$ for all $x\in \mathbb{R}$ this coordinate can take arbitrary values, see footnote \ref{foot}; $x$ resides in any connected interval where $P(x)>0$, and the boundary (or boundaries) of such interval, where $P(x)=0$, provides the possible values of $x$ where the axes (regular or not) corresponding to the cyclic $\varphi$ are placed; finally $q$ resides on any interval with a definite sign of $Q$ ---before any extension is performed---subject to the restriction that $q-\alpha x >0$.

According to \cite{OvcharenkoPodolskyAstorino:2024}, $m$~is the mass parameter, $a$~is a Kerr-like angular-momentum parameter, $l$~is a NUT parameter, $\alpha$~is an acceleration parameter, $e$~and $g$~are electric and magnetic charges, respectively, $\Lambda$~denotes the cosmological constant, and $C_f$ is a parameter governing the regularity or deficit angle of the possible axes of symmetry. For the physical dimensions of parameters and coordinates see \cite{OvcharenkoPodolskyAstorino:2024}. This seems as the natural interpretation for the outstanding known cases, but for cases with $\varphi\in \mathbb{R}$ and/or with $x$ with a semi-infinite range these names and units may need to be adapted.

\subsection{The range of the coordinate $x$}\label{subsec: P>0}
To identify the possible ranges of $x$, and as mentioned in the previous paragraph, we must identify the connected intervals of $x$ where the condition $P>0$ is kept. From its expression one readily checks that $P(x)$ is a quartic polynomial 
\[
P(x)=1+c_1x+c_2x^2+c_3x^3+c_4x^4,
\]
whose coefficients are
\begin{align*}
c_1&=-2\alpha m,\\
c_2&=\alpha^2(a^2+e^2+g^2-l^2)-1-\Lambda l^2,\\
c_3&=2\alpha m-\frac43\Lambda al,\\
c_4&=-\alpha^2(a^2+e^2+g^2-l^2)
-\frac{\Lambda}{3}\left[a^2+\alpha^2(a^2-l^2)^2\right].
\end{align*}
Therefore, the coefficients are (generically) {\em arbitrary} and independent except for the constant term $P(0)=1$. This implies that $x=0$ always belongs to a connected interval with $P>0$. Assuming that $m>0,\, \,  \alpha>0$ one also knows that $P'(0)=c_1=-2\alpha m<0$. Thus, the number of real roots of $P(x)$ and their signs depend basically on the sign of $c_4$. If $c_4>0$ then $P(\pm\infty) \rightarrow \infty$ and either (for simple roots)
\begin{itemize}
\item there are no real roots, and $P(x)>0$ on the whole real line. Then $x\in (-\infty,\infty)$ and there are no possible axes;\footnote{In this situation probably one should unwrap $\varphi$ and let its range be $\mathbb{R}$.\label{foot}} 
\item there are two positive real roots;
\item there are two negative real roots;
\item there are four different real roots.
\end{itemize}
Similarly, if $c_4<0$ there can be one negative and one positive root, or three negative ones and one positive, or one negative and three positive. Finally, there is the marginal case with $c_4 =0$, in which case $P(x)$ has degree three and the distribution of the real roots depends on the sign of $c_3$. The full classification --for the case of simple roots-- is given in \cref{table} ($x_i$, with $i=1,2,3,4$ denote the real roots in order if they exist).
\begin{table}[h!]
\centering
\begin{tabular}{lll}
\toprule
Condition & Real roots & Intervals where $P>0$\\
\midrule
$c_4>0$ & none & $\mathbb R$\\
& two positive & $(-\infty,x_1)\cup(x_2,\infty)$\\
& two negative & $(-\infty,x_1)\cup(x_2,\infty)$\\
& four & $(-\infty,x_1)\cup(x_2,x_3)\cup(x_4,\infty)$\\
\midrule
$c_4<0$ & one neg., one pos. & $(x_1,x_2)$\\
& three neg., one pos. & $(x_1,x_2)\cup(x_3,x_4)$\\
& one neg., three pos. & $(x_1,x_2)\cup(x_3,x_4)$\\
\midrule
$c_4=0,c_3>0$ & one negative & $(x_1,\infty)$\\
& two neg., one pos. & $(x_1,x_2)\cup(x_3,\infty)$\\
\midrule
$c_4=0,c_3<0$ & one positive & $(-\infty,x_1)$\\
& one neg., two pos. & $(-\infty,x_1)\cup(x_2,x_3)$\\
\midrule
$c_4=c_3=0,c_2>0$ & none & $\mathbb R$\\
& two positive & $(-\infty,x_1)\cup(x_2,\infty)$\\
\midrule
$c_4=c_3=0,c_2<0$ & one neg., one pos. & $(x_1,x_2)$\\
\midrule
$c_4=c_3=c_2=0$ & one positive & $(-\infty,x_1)$\\
\bottomrule
\end{tabular}
\caption{Distribution of simple real roots for the polynomial $P(x)$.}\label{table}
\end{table}

The cases involving roots of higher multiplicity can easily be derived from \cref{table}. Nevertheless, these are of little interest because the possible axes at roots of this kind are never regular, see next subsection.

\subsection{Study of the existence and regularity of symmetry axes}\label{subsec:cones}

\begin{table}[h!] \centering \small \setlength{\tabcolsep}{4pt} \renewcommand{\arraystretch}{1.15} \begin{tabular}{ |>{\raggedright\arraybackslash}p{1.5cm} |>{\raggedright\arraybackslash}p{4.7cm} |>{\raggedright\arraybackslash}p{4.7cm} |>{\raggedright\arraybackslash}p{4.0cm}| } \hline \textbf{Cases} & \textbf{Condition for unique axial Killing vector} & \textbf{Can both axis' components be regularised?} & \textbf{Conclusion} \\ \hline {Generic case} &  \[ \begin{aligned} &2l+a(x_i+x_j) [1-\alpha^2(l^2-a^2)]\\ &\quad -2l\alpha^2(l^2-a^2)x_ix_j=0 \end{aligned} \] & Not generically. Both components can be regular only if \[ A^2(x_i)P'^2(x_j) = A^2(x_j)P'^2(x_i) \] & Either both components of the axis are regular, or one of them has a conical deficit.\\ \hline $a=0$, $l=0$ & The condition is automatically satisfied; the axial Killing vector is \[ \partial_\varphi \] & No, except in the non-accelerating limit \[ \alpha=0 \] & For $\alpha\neq0$, at least one component must have a conical deficit. \\ \hline $l=0$, $\ a\neq0$ & \[ x_j=-x_i \] & No, except in the non-accelerating limit \[ \alpha=0 \]  & For $\alpha\neq0$, at least one component must have a conical deficit. \\ \hline $a=0$ , $l\neq0$ & \[ x_ix_j=\frac{1}{\alpha^2l^2} \] thus, the roots have the same sign & No & At least one component must have a conical deficit; $\alpha\neq 0$ is not possible. \\ \hline $\alpha=0$ & \[ x_i=-x_j-\frac{2l}{a} \] & Yes, if either \[l=0\] or if  \[\Lambda =3/l^2 \, \, \mbox{and} \, \, |a/l|>1 \] & For $l=0$, this is the Kerr--Newman--(A)dS family.

 {\bf These are the non-radiative cases}.
\\ \hline \end{tabular} \caption{Summary of the existence of an axial Killing vector and the regularity of its symmetry axis components.} \label{tab:symmetry-axis} \end{table}

To understand the existence of axes associated to the cyclic coordinate, and its regularity, we must consider an arbitrary Killing vector
$$
\partial_\varphi +\tau \partial_t
$$
where $\tau$ is a constant. To be axial, this Killing vector must vanish on the axis of symmetry \cite{MarsSenovilla1993}.
This requires that the Killing vector has vanishing norm and vanishing scalar product with any other vector. One can easily prove that this leads to the following conditions ---see \cref{tab:symmetry-axis} for a summary of the analysis---
\begin{equation}\label{axes}
P(x)=0, \qquad B(x) -\tau A(x) =0.
\end{equation}
Hence any possible axis of symmetry must lie at $x=x_i$, with $x_i$ a real root of $P(x)$. Once the root is known, the constant $\tau$ is fixed to be
$$
\tau_{x_i} = \frac{B(x_i)}{A(x_i)} = \frac{a+2l x_i +a x_i^2}{1+\alpha^2 (l^2-a^2) x_i^2} .
$$
The regularity of the axis (or its deficit/excess angle $\Delta$) is then defined by the limit \cite{MarsSenovilla1993}$$
\lim_{x\rightarrow x_i} \frac{g^{\mu\nu} \partial_\mu f \partial_\nu f}{4 f}=\left( 1-\frac{\Delta_{x_i}}{2\pi}\right)^2
$$
which must be one ($\Delta_{x_i} =0$) for regular axes, where $f$ is the norm square of the Killing vector. In our case, the left-hand side at any root $x=x_i$ of $P(x)$ can be computed straightforwardly and the result is
	\begin{equation}\label{eq:con}
	\brkt{\frac{1}{4C_f A^2(x)}\prn{\frac{\dd P(x)}{\dd x}}^2}\Bigg|_{{x=x_i}} =\left( 1-\frac{\Delta_{x_i}}{2\pi}\right)^2 \, \, .
	\end{equation}
It follows that double or higher-order roots cannot define a regular axis, and they are actually geometrical singularities\footnote{If $A(x_i)=0$ too, from \eqref{axes} $B(x_i)=0$ and the metric would be degenerate at $x_i$.}. 
However, at every simple root $x_i$ one can adjust $C_f$ such that there is no deficit angle. Thus, if one is dealing with the solution defined in one of the intervals of type $(x_i,\infty)$, or $(-\infty,x_i)$, the set $x=x_i$ is an axis of symmetry that can always be made regular by choosing $C_f$ appropriately, and the two-surfaces spanned by $\{q,\varphi\}$ may have a different topology ---for instance, if $c_4=c_3=0$ the topology will be $\mathbb{R}^2$. This depends on whether or not $\Omega^{-1} \rho \int^\infty du/\sqrt{P(u)}$ is finite at fixed values of $t,\varphi$ and $q$.

If, on the other hand, one is dealing with a solution with $x$ defined on a finite interval $(x_i,x_j)$, with $x_i$ and $x_j$ consecutive simple roots of $P(x)$ such that $P(x)>0$ on $(x_i,x_j)$, then either $\Delta_{x_i}$ or $\Delta_{x_j}$ can always be made to vanish by choosing $C_f$. However, the Killing vector for the other root will be different from the one regularized, because $\tau_{x_i}\neq \tau_{x_j}$ in general. If this is the case, one has to choose once and for all the axial Killing vector by defining the proper identification of points, with 
\begin{equation}\label{eq:identify}
\varphi \leftrightarrow \varphi +2 \pi \hspace{5mm} \mbox{ and {\em either}} \hspace{5mm}  t\leftrightarrow t+\tau_{x_i} 2 \pi \hspace{5mm} \mbox{ or} \hspace{5mm} t\leftrightarrow t+\tau_{x_j} 2 \pi \, .
\end{equation}
Once this choice is made, say with $\tau_{x_i}$, the chosen axis --via the choice of identification made in \eqref{eq:identify}--- can always be regularized by choosing the corresponding $C_f$ according to \eqref{eq:con}. However, the set of points $x=x_j$ where the other Killing vector vanishes is just its set of fixed points, but this is not an axis of symmetry any longer, as the orbits of $\partial_\varphi +\tau_{x_j}\partial_t$ are now open, diffeomorphic to $\mathbb{R}$, describing a sort of helices. As the radius of the helices decreases, their step also diminishes leading to the fixed points at $x=x_j$. Therefore, there is no question of regularizing $\partial_\varphi +\tau_{x_j}\partial_t$ at $x=x_j$, as this cannot be an axis of symmetry.\footnote{For a different interpretation, allowing for Killing vectors with open orbits to be called `axial', see \cite{KKO}.} In summary, in the generic case with $\tau_{x_i}\neq \tau_{x_j}$ there is one axial Killing vector, and the second Killing vector with fixed points is not circular unless both identifications in \eqref{eq:identify} are performed simultaneously. If this is done, the orbits of the group of symmetries are actually compact, and will lead to the existence of closed timelike curves (CTCs) if we are dealing with a region where $Q>0$. On the other hand, if $Q<0$ both Killing vectors are spacelike and the orbits can be either compact, or not, according to whether we make only one, or two, identifications from \eqref{eq:identify}. One must notice that, even if the second identification is not made, and we are in a region with $Q>0$, there will always be CTCs. This follows from the fact that   
$$
\partial_\varphi +\tau_{x_i} \partial_t =  \left(\partial_\varphi +\tau_{x_j} \partial_t \right) +(\tau_{x_i} -\tau_{x_j}) \partial_t
$$
so that around the axis of symmetry of the chosen axial Killing vector, the other Killing vector is proportional to $\partial_t$ and thus timelike if $Q>0$ ---unless $\tau_{x_i} -\tau_{x_j}=0$.

Hence, a more favorable case arises when the Killing vector that vanishes at $x_i$ and $x_j$ is the same. The condition to have the same Killing vector with fixed points on $x=x_i$ and $x=x_j$ is obviously $\tau_{x_i}=\tau_{x_j}$, and this requires the following condition
$$
A(x_i) B(x_j) - A(x_j) B(x_i)=0
$$
which, given that ($x_i-x_j)$ always factorizes, simplifies on using \eqref{eq:A}--\eqref{eq:B} to 
\begin{equation}\label{eq:sameK}
2l+a (x_i+x_j) [1-\alpha^{2}(l^{2}-a^{2})]-2l \alpha^2 (l^2-a^2) x_i x_j=0.
\end{equation} 
 Trivially, if $a=l=0$ \eqref{eq:sameK} is an empty condition, as it is obvious because then any $\tau_{x_i}$ vanishes. However, interesting results can be derived for other cases. To start with, if $l=0$ then \eqref{eq:sameK} reduces to 
\begin{equation}\label{eq:sameK1}
a (x_i+x_j)\left[ 1 +\alpha^{2}a^{2}\right]=0
\end{equation} 
implying (with $a\neq 0$) that the two roots are placed symmetrically with respect to $x=0$: $x_j=-x_i$. If on the other hand $a=0$ but $l\neq 0$, then \eqref{eq:sameK} collapses to
$$
x_i x_j =\frac{1}{\alpha^2 l^2}
$$
so that both roots must have the same sign ---again assuming that $\alpha\neq 0$. If $\alpha =0$ then the condition is simply
$$
2l+a(x_i+x_j)=0.
$$

When \eqref{eq:sameK}  is satisfied the unique axial Killing vector has an axis (its set of fixed points) with {\em two} disconnected components. Regularization can always be achieved at any one of these two connected components. Whether or not the second one can also be regularized without deficit angle is considered next.

The condition for having the same deficit/excess angle at both roots $x_i,x_j$, in which case one can always further choose $C_f$ so that the axis is regular at both ends is ($'=d/dx$)
$$
A^2(x_i) P'^2(x_j) =A^2(x_j) P'^2(x_i)
$$
according to \eqref{eq:con}. If the above condition is not satisfied, then one of the components of the axis can always be regularized, the other will have a deficit/excess angle. 

As a consequence, there is a unique Killing vector with no deficit angle at both $x=x_i$ and $x=x_j$, with $x_i$ and $x_j$ finite simple real roots of $P(x)$ if, and only if,
\begin{equation}\label{Fullcondition}
\frac{A(x_i)}{A(x_j)} =\frac{B(x_i)}{B(x_j)} =\epsilon \frac{P'(x_i)}{P'(x_j)}
\end{equation}
with $\epsilon^2 =1$, and both signs must be taken into consideration. From our hypothesis, 
$$
P(x)=(x-x_i)(x-x_j) R(x) 
$$
where $R(x)<0$ on $[x_i,x_j]$ ($x_i < x_j$). Then, 
\begin{equation}\label{eq:R}
\frac{P'(x_i)}{P'(x_j)} =- \frac{R(x_i)}{R(x_j)}<0 .
\end{equation}

\subsubsection{Case $a=0=l$}
In this situation the axial Killing is $\partial_\varphi$. Now we have $c_4<0$ and thus either there are 2 or 4 roots. Two of them are always $x=\pm 1$ and the other two exist if and only if 
$$
m^2 > e^2 +g^2
$$
and are given by
$$
x_\pm = \frac{1}{\alpha (e^2 +g^2)}\left( m \pm \sqrt{m^2 -e^2 -g^2}\right)
$$
both of them are positive.
When these two roots do exist, there are still several possibilities for the connected open interval with $P(x)>0$
\begin{itemize}
\item if $m - \sqrt{m^2 -e^2 -g^2} >\alpha (e^2 +g^2)$ then the interval for $x$ is either $(-1,1)$ or $(x_-,x_+)$;
\item if $m + \sqrt{m^2 -e^2 -g^2} <\alpha (e^2 +g^2)$ the possible intervals are $(-1,x_-)$ and $(x_+,1)$;
\item if $m - \sqrt{m^2 -e^2 -g^2} <\alpha (e^2 +g^2)$ and $m + \sqrt{m^2 -e^2 -g^2} >\alpha (e^2 +g^2)$ the intervals are $(-1,x_-)$ and $(1,x_+)$.
\end{itemize}

Condition \eqref{eq:sameK} is empty and now $A=1$, thus \eqref{Fullcondition} reduces to the quotient of $P'$s on the extremes of the interval to be $\pm 1$. For the interval $(-1,1)$, this reduces, on using \eqref{eq:R} and the values of $x_++x_-$ and $x_+ x_-$,  to
$$
\frac{P'(-1)}{P'(1)} = -\frac{\alpha^2(e^2+g^2) +2\alpha m +1}{\alpha^2(e^2+g^2) -2\alpha m +1} =\pm 1.
$$
This is impossible for the $-$ sign, and for the $+$ sign implies $\alpha m=0$ and (assuming $m\neq 0$) this leads to the case $\alpha =0$ (to be considered later in full generality).

Again using \eqref{eq:R} one can easily compute 
$$
P'(x_\pm) = \pm \alpha^2 (e^2+g^2) (1-x_\pm^2) (x_+ -x_-)
$$
and thus
$$
\frac{P'(x_-)}{P'(x_+)} = -\frac{(e^2+g^2)[\alpha^2(e^2+g^2)+1] -2m^2 +2m \sqrt{m^2 -e^2 -g^2}}{(e^2+g^2)[\alpha^2(e^2+g^2)+1] -2m^2 -2m \sqrt{m^2 -e^2 -g^2}}=\pm 1
$$
implies, for $m\neq 0$, that $(e^2+g^2)[\alpha^2(e^2+g^2)+1] = 2m^2$ and, {\em a fortiori}, $1< \alpha^2 (e^2+g^2)$ which happens to contradict the needed condition $m - \sqrt{m^2 -e^2 -g^2} >\alpha (e^2 +g^2)$ allowing for the $(x_-,x_+)$ interval.  

The next possibility is $(-1,x_-)$, and using the previous computations for the values of $P'(-1)$ and $P'(x_-)$ this leads to
$$
\pm \alpha(e^2 +g^2) \left[ \alpha (e^2+g^2) +m +\sqrt{m^2 -e^2 -g^2}\right] =\sqrt{m^2 -e^2 -g^2}\left[m-\sqrt{m^2 -e^2 -g^2}-\alpha(e^2+g^2)\right]
$$
which can be satisfied only for the $+$ sign leading to
$$
\alpha(e^2 +g^2) \left[ \alpha (e^2+g^2) +m +2\sqrt{m^2 -e^2 -g^2}\right] =\sqrt{m^2 -e^2 -g^2}\left(m-\sqrt{m^2 -e^2 -g^2}\right).
$$
However, this implies $\alpha(e^2+g^2) < m - \sqrt{m^2 -e^2 -g^2}$ which is not compatible with the conditions for the existence of the interval $(-1,x_-)$. 

The remaining possibility is $x\in (1,x_+)$ or $x\in (x_+,1)$, as the calculation is identical for both cases. A similar computation using \eqref{eq:R} provides
$$
-\frac{P'(x_+)}{P'(1)} = \frac{(1+x_+)(x_+-x_-)}{1-x_-}=\pm 1
$$
leading for each sign to either
$$
\left[ \alpha(e^2+g^2) -m -\sqrt{m^2 -e^2 -g^2}\right] \left[\alpha(e^2+g^2)+\sqrt{m^2 -e^2 -g^2} \right]=0
$$
or
$$
\alpha(e^2+g^2) \left[ \alpha(e^2+g^2) -m +2\sqrt{m^2 -e^2 -g^2} \right]=-\sqrt{m^2 -e^2 -g^2} (m+\sqrt{m^2 -e^2 -g^2} ).
$$
The former expression leads to $x_+=1$ and must be discarded. The latter easily implies that $\alpha(e^2+g^2) > m+\sqrt{m^2 -e^2 -g^2}$ that enters in contradiction with the requirements to have either $(1,x_+)$ or $(x_+,1)$. 

\subsubsection{Case $l=0$ (with $a\neq 0$)}
As proven above, this case must have $x_i=-x_j$ to keep a unique axial Killing vector. Now, setting $P(x_i)\pm P(-x_i)=0$ one easily derives that the only possibilities are given by $\alpha=0$ ---to be considered later--- or by $x_j=1=-x_i$ with $\Lambda =0$. For the latter, one has $A(\pm 1)=1-\alpha^2 a^2$ and thus the condition for unique $C_f$ reads
$$
\frac{1-2\alpha m +\alpha^2(a^2 +e^2 +g^2)}{1+2\alpha m +\alpha^2(a^2 +e^2 +g^2)}=\pm 1
$$
and again this can only happen when $\alpha =0$. 

\subsubsection{Case $a=0$ (with $l\neq 0$)}
From above we know that $x_i x_j =1/(\alpha^2 l^2)$ and both of them must have the same sign. Now $A(x) =1+\alpha^2 l^2 x^2$ and thus the condition to have unique Killing with same $C_f$ becomes
$$
\epsilon \frac{R(x_i)}{R(x_j)} = \frac{1+\alpha^2 l^2 x_i^2}{1+\alpha^2 l^2 x_j^2}=\frac{x_i}{x_j}
$$
and letting $R(x)=K(1+\gamma x +\delta x^2)$ this leads to
$$
x_j(1+\gamma x_i +\delta x_i^2)=\epsilon x_i (1+\gamma x_j+\delta x_j^2)
$$
which using again $x_i x_j =1/(\alpha^2 l^2)$ readily implies $x_i =\epsilon x_j$ which is impossible for simple roots of the same sign.

\subsubsection{Case $\alpha=0$}
For the unique Killing we proved that $x_i=-x_j-2l/a$ and now $P(x) =1-x^2 -\frac{\Lambda}{3}x^2(3l^2 +4al x +a^2 x^2)$ and $A(x)=1$. Then 
$$
P(x)-P(-x-2l/a) = \frac{4l}{a^2} \left(1-\frac{\Lambda}{3} l^2\right) (ax+l)
$$
hence either $x_i=x_j =-l/a$ leading to a double root, or $\Lambda=3/l^2 >0$ or $l=0$. The latter case leads to the Kerr-Newman-(A)dS solution. On the other hand, when $\Lambda =3/l^2$ the roots are $(l/a)(-1\pm \sqrt{1+a/l)})$ and $(l/a)(-1\pm \sqrt{1-a/l})$ and the quotient $P'(x_i)/P'(x_j)$ between {\em consecutive} roots enclosing an interval with $P(x)>0$ can be $\pm 1$ only if
$$
\left|\frac{a}{l}\right|>1 .
$$

The results to be proven  in this work show that $\alpha=0$ if and only if there is no gravitational radiation ---see \cref{sec:radiation-Lambda-p,sec:radiation-Lambda-n}. This reveals a relation between the presence of a deficit/excess angle and the presence of gravitational radiation  at infinity using the asymptotic super-Poynting vector \cite{Fernandez-Senovilla2020a} (see \cref{eq:s-P-scri,eq:s-Poynting-n} for  $\Lambda\neq 0$, noting that one can do the  limit of the $\Lambda>0$ formula to $\Lambda=0$ \cite{Fernandez-Senovilla:2022b}): for the case of a unique well-defined axial Killing vector, $\alpha\neq 0$ if and only if one of the components of the axis presents a non-zero deficit/angle, and if and only if there is gravitational radiation at infinity.

\section{Conformal space-time and infinity}
\label{sec:conformal-spacetime}

The unphysical metric ${g_{\alpha\beta} = \Omega^2\,\hat{g}_{\alpha\beta}}$ corresponds to the conformal compactification $(M,\ct{g}{_{\alpha\beta}})$ of the physical space-time $(\hat{M},\ct{\hat{g}}{_{\alpha\beta}})$ with  metric given by \cref{A-new-metric}. Such conformal completions \cite{Penrose63,Kroon2016} require $\Omega>0$ in $M\setminus\scri$, while $\Omega=0$ at the boundary $\scri$ of the conformal manifold $M$. Thus, in view of \cref{Omega_new}, one has
	\begin{align}
		\Omega=q -\alpha x &> 0 \quad\text{ at points of } M\setminus\scri\ ,\label{eq:condition-bulk}\\
	 \Omega= q -\alpha x &= 0 \quad\text{ at points on }  \scri\ . \label{Omega=0}
	\end{align}
The boundary $\scri$ is a differential manifold itself that represents infinity, and it is the natural setting in which to study gravitational radiation \cite{Penrose65,Geroch1977} ---for a review on conformal infinity see \cite{Frauendiener2004}. There is a gauge freedom $\Omega \rightarrow \Omega \omega$ for any $\omega >0$ in the choice of the conformal factor, and thus only the conformal structure of $\scri$ is fixed.

With this compactification, the conformal metric can be read from the terms between brackets in \cref{A-new-metric}. The one form	
\begin{equation}
N_\beta \defeq \nabla_\beta \Omega=\partial_{\beta}q-\alpha\partial_{\beta}x\ \label{eq:normal-scri}
\end{equation}
is normal to $\scri$. If we raise its index with the conformal metric, we get
\begin{equation}\label{eq:normal-scri-explicit}
 \ct{N}{^{\alpha}} = \frac{1}{{\rho}^2}\prn{
     {Q} \delta_{q}^\alpha - \alpha\,{P}\delta_{x}^\alpha}\ ,
\end{equation}
and, in agreement with the conformal Einstein Field Equations ---see, e.g., \cite{Paetz2013}---, on $ \scri $ we get
	\begin{equation}
	{N}_{\mu}{N}^{\mu}\Big|_\scri =\frac{1}{\rho^2}\prn{{Q}+\alpha^2{P}}\Big|_\scri= -\frac{\Lambda}{3} \, . \label{eq:efe-scri}
	\end{equation}
Notice that the causal character of $\ct{N}{^{\alpha}}$ is, as it should be, null for $\Lambda=0$, timelike for $\Lambda>0$, and spacelike for $\Lambda<0$. For the cases $\Lambda\neq 0$, at $\scri$ we can define the unit normal 
\begin{equation}	\label{eq:unit-normal-scri}
\ct{n}{^{\alpha}}\defeq \sqrt{\frac{3}{|\Lambda|}}\ct{N}{^{\alpha}}\ .
\end{equation}

Equation \eqref{eq:efe-scri} shows that $Q|_\scri <0$ if $\Lambda>0$, and both signs of $Q|_\scri$ can occur when $\Lambda<0$. Also, if $\alpha =0$ and $\Lambda <0$ then $Q|_\scri$ is necessarily positive. Put another way, $Q|_\scri < 0$ requires $\alpha\neq 0$ if $\Lambda <0$. However, noticing that
$$
\rho^2|_\scri = A^2 +\alpha^2 B^2 |_\scri = \left(1+\alpha^2(l^2-a^2)x^2\right)^2 +\alpha^2(a+2lx+ax^2)^2 >0
$$
\eqref{eq:efe-scri} shows that, around any value $x=x_i$ with $P(x_i)=0$, $Q|_\scri (x_i)>0$ if $\Lambda$ is negative. Hence, for $\Lambda <0$ one of the following possibilities must happen ---see \cref{table:Q-boundary-results}---
\begin{itemize}
\item $Q|_\scri >0$ for the entire allowed range of $x$. From 
\begin{equation}\label{eq:Q(0)}
Q|_\scri (0)=-\alpha^2 -\frac{\Lambda}{3}(1+ \alpha^2 a^2)
\end{equation}
we obtain a necessary condition for this to happen whenever $x=0$ belongs to the interval in consideration:
\begin{equation}\label{eq:large}
-\frac{\Lambda}{3} >\frac{\alpha^2}{1+\alpha^2 a^2}.
\end{equation}
This requires
$$
-\frac{\Lambda}{3} > \frac{1}{a^2} \hspace{3mm} \mbox{if} \hspace{3mm}\alpha\neq 0, \hspace{3mm} \mbox{and} \hspace{3mm} -\frac{\Lambda}{3} >\alpha^2 \hspace{3mm}\mbox{if} \hspace{3mm} a=0.
$$
For a general connected  interval $I$ with $P(x)>0$ a necessary condition reads
\begin{equation}\label{eq:Large}
-\frac{\Lambda}{3} >\alpha^2 \max_{I} \frac{P(x)}{\rho^2_\scri} .
\end{equation}
 In case that the range for $x$ is (semi)-infinite, another necessary condition is derived from the behaviour at $x\rightarrow \pm \infty$, and reads
$$
l^2 < a^2 +e^2 +g^2 .
$$
\item $Q|_\scri$ vanishes at two (or four) values of $x$ within the allowed range for $x$. This can happen when the allowed range is a finite interval $(x_i,x_j)$ or a semi-infinite interval ($c_4 \geq 0$, see table \ref{table}). In this case, one must only consider one of the intervals with $Q|_\scri >0$, or one of the connected intervals with $Q|_\scri <0$ (and thus $\alpha\neq 0$). To be able to say something including points where $Q=0$, one should first perform an appropriate extension of the metric\footnote{Notice that if such an extension can be performed, only the case $\alpha\neq 0$ matters even at the region with $Q|_\scri >0$, because $\alpha\neq 0$ is needed to have a portion with $Q|_\scri <0$.}. As we will see later on, actually the places with $Q|_\scri =0$ will not be part of the conformal metric at $\scri$ --and thus, they must be excluded.
\item $Q|_\scri <0$ for the entire allowed range of $x$. Then $\alpha\neq 0$ and from \eqref{eq:Q(0)} a necessary condition is
$$
-\Lambda/3 <\alpha^2 /(1+\alpha^2 a^2).
$$
This situation is only possible if $P(x)$ has no zeros ($c_4 >0$ or $c_4=c_3 =0$ with $c_2 >0$), and hence there is no axial symmetry.  Moreover, this case requires $l^2 > a^2 +e^2 +g^2$ as otherwise $Q|_\scri$ would diverge to $+\infty$ for large values of $x$.
\end{itemize}

\begin{table}[ht!] \centering \small \setlength{\tabcolsep}{4pt} \renewcommand{\arraystretch}{1.15} \begin{tabular}{ |>{\raggedright\arraybackslash}p{2.8cm} |>{\raggedright\arraybackslash}p{5.2cm} |>{\raggedright\arraybackslash}p{6.4cm}| } \hline \textbf{Sign of $Q|_{\mathscr J}$} & \textbf{Conditions / necessary conditions} & \textbf{Consequences}   \\ \hline \boldmath$Q|_{\mathscr J}>0$ throughout the allowed $x$--range & If $0\in I$, \[ -\frac{\Lambda}{3} > \frac{\alpha^2}{1+\alpha^2a^2}. \] More generally, \[ -\frac{\Lambda}{3} > \alpha^2 \max_I \frac{P(x)}{\rho_{\mathscr J}^2}. \] For a (semi-)infinite range, \[ l^2<a^2+e^2+g^2. \] &  $Q$ remains positive throughout the selected connected range, including at any symmetry-axis points $P(x_i)=0$. \\ \hline \boldmath$Q|_{\mathscr J}=0$ at two or four points & Possible for finite or semi-infinite allowed $x$--ranges. & The connected regions where $Q>0$ and $Q<0$ must be considered separately.  The points where \[ Q|_{\mathscr J}=0 \] do not belong to the conformal metric at $\mathscr J$ and must be excluded. \\ \hline \boldmath$Q|_{\mathscr J}<0$ throughout the allowed $x$--range & Requires $\alpha \neq 0$ and \[ -\frac{\Lambda}{3} < \frac{\alpha^2}{1+\alpha^2a^2}. \] In addition, \[ P(x)\neq0 \] throughout the range. & There are no roots of $P$, hence no symmetry axis. For a complete (semi-)infinite range one also needs \[ l^2>a^2+e^2+g^2. \]  An everywhere-negative $Q|_{\mathscr J}$ is possible only in the absence of axial symmetry. \\ \hline  \end{tabular} \caption{Summary of possibilities concerning the sign of $Q|_{\mathscr J}$  for $ \Lambda<0 $. Notice that $Q_\scri <0$ when $\Lambda >0$.} \label{table:Q-boundary-results} \end{table}

Next,  we can introduce coordinates $y^a=\prn{\bar{t},\bar{x},\bar{\varphi}}$ in $\scri$, and doing the natural identification $x(y^a)=\bar{x}$, $q(y^a)=\alpha\bar{x}$, $t=\bar{t}$, $\varphi=\bar{\varphi}$, take the pullback of the space-time metric to $\scri$ in order to get a representative of the conformal metric at $\scri$. This gives
\begin{equation}
h =
    \dfrac{\bar{P}}{C_f\bar{\rho}^2}\,(\omega^1)^2
   -\dfrac{\bar{Q}}{C_f\bar{\rho}^2}\,(\omega^2)^2
   - \dfrac{\Lambda}{3}\,\frac{\bar{\rho}^4}{\bar{P} \bar{Q}}\,(\omega^3)^2
   \,,
    \label{conformal-metric-in-cobasis}
\end{equation}
where the one-form co-basis is
\begin{equation}
\omega^1 \defeq \bar{A}\dd\bar{\varphi} + \alpha^2 \bar{B}\dd \bar{t}\ ,\qquad
\omega^2 \defeq \bar{A}\dd \bar{t} - \bar{B}\dd\bar{\varphi}\ ,\qquad
\omega^3 \defeq \dd \bar{x}\ ,
    \label{cobasis-on-scri}
\end{equation}
and barred functions denote the pullback to $\scri$ of \cref{eq:A} -- \cref{eq:P-hom} \footnote{Not to be confused with the complex conjugation of the rescaled Weyl scalar $\phi_2$ and complex vector field $\ct{{m}}{^\alpha}$ of a null tetrad introduced later on.}.
From this form, the determinant can be easily computed,
\begin{equation}
\det h= \frac{\Lambda}{3C_f^2}\ , 
\end{equation}
showing that it has the correct sign depending on $\Lambda$: if ${\Lambda>0}$, the metric \eqref{conformal-metric-in-cobasis} is Riemannian with signature ${(+,+,+)}$; if ${\Lambda<0}$, it is Lorentzian with ${(+,-,+)}$ ($\bar{Q}>0$) or ${(+,+,-)}$ ($\bar{Q}<0$); and, if $\Lambda=0$, the metric is degenerate.

	\subsection{Conformal Killing vector fields at $\scri$}\label{sec:kvfs}
For the particular representative \eqref{conformal-metric-in-cobasis} of the conformal structure at $\scri$, consider the general KVF in \eqref{eq:generalKVF} with arbitrary constants $\omega$ and $\tau$.
For the general conformal structure, these are actually conformal Killing vector fields (CKVFs). Observe that these vector fields become tangent to $\scri$, as they are orthogonal to the normal \eqref{eq:normal-scri}. Of course, they are always spacelike for $\Lambda >0$. Thus, let us focus for the rest of this section on the cases with negative $\Lambda<0$. For $\Lambda <0$ their causal character at infinity  has implications in later analysis, and for this purpose we present here two alternative expressions of their norm, in terms of the metric functions at $\scri$ (all expressions are evaluated at $\scri$):
	\begin{equation}
	    \ct{\xi}{^\mu}\ct{\xi}{_\mu} =   \frac{1}{C_f\rho^2}\brkt{P\prn{\tau A+ \omega \alpha^2 B}^2-Q\prn{\omega A- \tau B}^2}\ .\label{eq:normKVFs-1}
	\end{equation}
 In addition, using \eqref{eq:efe-scri}, \cref{eq:normKVFs-1} can be recast into
	\begin{equation}
		\ct{\xi}{^\mu}\ct{\xi}{_\mu} =   \frac{1}{C_f}\brkt{\prn{\tau^2+\alpha^2\omega^2}P+\frac{\Lambda}{3} \prn{\omega A- \tau B}^2}\ .\label{eq:normKVFs-2}
	\end{equation}
We see that at points where $Q<0$ (so that $\alpha\neq 0$), 
there are no timelike CKVFs at $\scri$, hence these regions can therefore be regarded as non-conformally-stationary. However, for $Q>0$ there can be timelike CKVFs tangent to $\scri$. Actually, they always exist near the roots $x_i$ of $P(x)$, as follows from \eqref{eq:normKVFs-2}.
Observe also that if $Q$ has zeros (i.e. we are in the second-bullet situation of the previous page), then sufficiently close to those points these CKVFs become spacelike, that is, they cannot be timelike everywhere on the entire $\scri$, and therefore it is not a \emph{globally} conformally stationary region.
It remains the case with $Q|_\scri >0$ everywhere on the allowed range of $x$. This requires a {\em sufficiently negative} $\Lambda$ as shown in \eqref{eq:large}.  Such a region of $\scri$ may be globally conformally stationary in the sense of containing at least one CKVF that is  timelike everywhere.

As an example, consider $l=0$, $a=0$ with $\alpha\neq 0$. This makes $B=0$, and it is then clear that $\tau=0$ in \cref{eq:generalKVF,eq:normKVFs-1} gives a KVF with negative norm. Also, $\rho=1$ in this case, so that by \cref{eq:efe-scri}, the zeros of $Q$ would be placed at $3\alpha^2 P=-\Lambda$. If in the considered region $P$ is bounded (see \cref{table}), i.e., we are between 2 roots of $P$, for sufficiently negative values of $\Lambda$ no zeros of $Q$ will occur. For those values, then, the region will be globally conformally stationary. On the other hand, if we are in a region with $P$ unbounded, regardless of how negative $\Lambda$ is, $Q|_\scri$ will approach a vanishing value at which $\ct{\xi}{^{\mu}}\ct{\xi}{_{\mu}}\rightarrow 0$ and $\scri$ will cease to exist, and near this end global conformal stationarity is broken. 

Apart from that, if we consider $\alpha=0$ (keeping the rest of parameters arbitrary), \cref{eq:efe-scri} gives $Q>0$ everywhere for $\Lambda<0$, so that $\scri$ is globally conformally stationary. Hence, a key point when $\Lambda<0$ is that cases with $Q|_\scri >0$ and  $P$  defined on a bounded interval are conformally stationary when $\alpha\neq 0$ for sufficiently negative values of $\Lambda$ --as in \eqref{eq:large} and \eqref{eq:Large}. This seems to indicate that a `purely' AdS behaviour dominates near $\scri$ when $|\Lambda|/3$ is large compared to $\alpha^2$, even in cases where there is radiation reaching $\scri$. 

\subsection{Principal null directions and rescaled Weyl scalars}
We adapt the principal null directions presented in \cite{Podolsky-toappear} to the conformal space-time and infinity. For $Q<0$ they read
\begin{align}
	\ct{k}{^\alpha} &= -\frac{1}{\sqrt{2}\rho} \brkt{ \frac{\sqrt{C_f}}{\sqrt{-Q}}
	 \prn{D\delta^\alpha_{t}-C\delta^\alpha_{\varphi}} + \sqrt{-Q} \delta^\alpha_q }\ , \nonumber \\
	\ct{l}{^\alpha} &= \frac{1}{\sqrt{2}\rho}\ \brkt{ \frac{\sqrt{C_f}}{\sqrt{-Q}}
	 \prn{D\delta^\alpha_{t}-C\delta^\alpha_{\varphi}} - \sqrt{-Q} \delta^\alpha_q } \ ,   \label{nullframe-Q<0}
\end{align}
whereas for $Q>0$ one has,
\begin{align}
	\ct{k}{^\alpha} &= \frac{1}{\sqrt{2}\rho} \brkt{ \frac{\sqrt{C_f}}{\sqrt{Q}}
	 \prn{D\delta^\alpha_{t}-C\delta^\alpha_{\varphi}} - \sqrt{Q}  \delta^\alpha_q } \ , \nonumber \\
	\ct{l}{^\alpha} &= \frac{1}{\sqrt{2}\rho}\ \brkt{ \frac{\sqrt{C_f}}{\sqrt{Q}}
	 \prn{D\delta^\alpha_{t}-C\delta^\alpha_{\varphi}} + \sqrt{Q} \delta^\alpha_q} \ ,  \label{nullframe-Q>0}
\end{align}
They have been normalised so that they form a pair of null tetrads once they are complemented with the complex vector field
\begin{equation}
	\ct{m}{^{\alpha}} =  \frac{1}{\sqrt{2}\rho} \brkt{
	 \frac{\sqrt{C_f}}{\sqrt{P}} \prn{B\delta_{t}^\alpha+A\delta_{\varphi}^\alpha} - \mathrm{i}  \sqrt{P} \delta^\alpha_x }\ .
\end{equation}

Using these bases it is possible to compute the Weyl scalars. As expected, only $\psi_2$ survives,
	\begin{equation}
		\psi_0=0\ ,\quad\psi_1=0\ ,\quad \psi_2\neq 0\ , \quad \psi_3=0\ ,\quad \psi_4=0\ ,
	\end{equation}
and $ \psi_2=0 $ at $\scri$ (the Weyl tensor vanishes there). As usual, one defines the rescaled Weyl tensor
	\begin{equation}\label{eq:rescaled-Weyl}
		\ct{d}{_{\alpha\beta\gamma}^{\delta}}\defeq \frac{1}{\Omega}\ct{C}{_{\alpha\beta\gamma}^{\delta}}\ .
	\end{equation}
The rescaled Weyl scalar $\phi_{2}$ can be evaluated at $\scri$ yielding the following finite expression:

	\begin{align}
	\phi_{2}&= -\frac{1}{(A-\im\,\alpha B)^3}
	      \bcbrkt{m\brkt{1+\alpha^2(l^2-a^2)-2\im\, \alpha a} 
	         -\im\, l\brkt{1-\alpha^2(l^2-a^2)} }\nonumber\\
	&  +\frac{\alpha (e^2+g^2)}{(A^2+\alpha^2B^2)
	   (A-\im\,\alpha B)^2}
	 \bcbrkt{2(1-\alpha^2a^2)x - \alpha^2 a l (1+x^2) \nonumber\\
		& - \im\,\alpha\,l\,[2+\alpha^2(a^2-l^2)]x^2
	       -\im\,\alpha\,(4 a\, x+ l)}\nonumber\\
	& -\frac{\Lambda}{3}\frac{\im\,l^3}{(A-\im\,\alpha B)^3}\, . \label{phi2}
			\end{align}
	Observe that the dependence on the mass $m$ is only in the first term, the dependence on the charges $e, g$ is in the second one, while the dependence on $\Lambda$ is in the last one. Some special subcases at $\scri$ are:
	
	\vspace{5mm}

	${a=0=l}$ and ${e=0=g}$:
	\begin{equation}
		\phi_{2} = -m \,.
	\end{equation}

	${a=0=l}$:
	\begin{equation}
		\phi_{2}= -m + 2\alpha\, (e^2+g^2)x\ .
	\end{equation}
	
	${l=0}$ and ${e=0=g}$:
	\begin{equation}
	  \phi_{2}= -\frac{m}{\prn{a\alpha+i}\prn{a\alpha x^2+i}^3}\,.
	\end{equation}

	${l=0}$:
	\begin{equation}
	  \phi_{2}=\frac{ -m - \im\,\alpha a\,m\,[\,1+(1 +\im\,\alpha a)\, x^2]+2\alpha\, x\,(e^2+g^2)}
	{(1+a^2\alpha^2)(1 + \im\, \alpha a\,x^2)(1 - \im\,\alpha a\, x^2 )^3}\,.
	\end{equation}

	${a=0}$ and ${e=0=g}$:
			\begin{equation}
				\phi_{2}=-\frac{1}{(1+\alpha^2 l^2x-\im\,\alpha l x)^3}\cbrkt{m\,(1+\alpha^2 l) 
	   - \im\,l\,\big[1-(\alpha^2+\frac{1}{3}\Lambda)\,l^2\big]}\,.
			\end{equation}

\section{Gravitational radiation with ${\Lambda>0}$}
\label{sec:radiation-Lambda-p}
	
The presence of gravitational radiation with $\Lambda>0$ was studied in \cite{FernandezPodolskySenovilla:2024} with a previous form of the metric \cite{PodolskyVratny2021,PodolskyVratny2023}. Unfortunately, that form of the metric could not describe the whole family of spacetimes and its special cases because it presented a degeneracy on the parameters $\alpha$ and $a$, in the sense that the vanishing of $a$ automatically implied the disappearance of $\alpha$. In particular, accelerating purely NUT black holes were not included in the study. For that reason, we briefly reconsider that analysis in this section in order to make it fully complete.

For the metric \eqref{A-new-metric} one can readily see that the two PNDs in \cref{nullframe-Q>0} are coplanar with the normal to $\scri$ \eqref{eq:normal-scri-explicit} if and only if ${\alpha=0}$. This implies that \emph{spacetimes contain gravitational radiation arriving at $\scri$ if and only if the acceleration parameter is non-zero}, as follows from Remark IV.4 in  the general study \cite{Fernandez-Senovilla:2022b}.\\

In fact, the above geometrical condition follows from the asymptotic super-Poynting $\ct{\overline{\P}}{^a}$ criterion \cite{Fernandez-Alvarez-Senovilla2020b,Fernandez-Senovilla:2022b}, built upon the rescaled Weyl tensor $\ct{d}{_{\alpha\beta\gamma}^{\delta}}$. For the present case, one can use the simplified formula for type-D spacetimes presented in \cite{FernandezPodolskySenovilla:2024}. Such expression is computed using \cref{nullframe-Q<0,eq:unit-normal-scri} and the coefficients\footnote{Observe that $c=0$ if and only if $\alpha=0$.}
\begin{equation}
 c\defeq \sqrt{2}\ct{\bar{m}}{^\mu}\ct{n}{_{\mu}}      = - \alpha\,\,\im\,\sqrt{\frac{3}{\Lambda}}\, \frac{\sqrt{P}}{\rho} \ ,\quad
 b^{-2} =-\sqrt{2}\ct{k}{^\mu}\ct{n}{_{\mu}} \sqrt{\frac{3}{\Lambda}}\, \frac{\sqrt{-Q}}{\rho} 
    \label{c-and-b}
\end{equation}
and reads in space-time coordinates
\begin{equation}\label{eq:s-P}
 \ct{\overline{\P}}{^\alpha} = 18\ \alpha \prn{\frac{3}{\Lambda}}^{\frac{5}{2}}
    \frac{PQ}{\rho^6 }\prn{Q-\alpha^2P}\phi_2\bar\phi_2
     \prn{\delta^\alpha_x + \alpha \delta^\alpha_q}\ .
\end{equation}
The asymptotic super-Poynting is tangent to $\scri$ (i.e., orthogonal to $\ct{N}{_{\alpha}}$), and its intrinsic expression on $\scri$ has all the information contained in the space-time version \eqref{eq:s-P}: 
\begin{equation}\label{eq:s-P-scri}
\ct{\overline{\P}}{^a} = \overline{\P}^{\bar{x}} \delta^a_{\bar x}\ ,
\end{equation}
with
\begin{align}
\overline{\P}^{\bar{x}}(\bar{x}) = 18 \alpha \prn{\frac{3}{\Lambda}}^{\frac{5}{2}}
    \frac{\bar{P}\bar{Q}}{\bar{\rho}^6 }\prn{\bar{Q}-\alpha^2 \bar{P}}\phi_2\bar\phi_2\ .
\end{align}
From this equation one deduces again the main conclusion just mentioned above,
\begin{center}
\emph{there is gravitational radiation at $\scri$ if and only if the acceleration parameter is non-zero.}
\end{center}
For completeness, we compute the asymptotic super-energy density, which is the component along $\ct{n}{^\alpha}$ of the full asymptotic supermomentum  \cite{Fernandez-Alvarez-Senovilla2020b} (all functions are evaluated at $\scri$):
\begin{equation}
\mathcal{W}=6\phi_2\bar{\phi}_{2} \brkt{1-6\,\alpha^2\Big(\frac{3}{\Lambda}\Big)^2\,
    \frac{{P} {Q}}{{\rho}^4 }} .
\label{super-energy-scri}
\end{equation}
For vanishing acceleration, $\alpha$, $\mathcal{W}$ is constant, and for $e=0=g=l=a=\alpha$ simply reads ${\mathcal{W}=6m^2} $.\\

		\subsection{The Cotton-York and the initial data tensor $\ct{D}{_{\alpha\beta}}$}\label{sec:C-D-dS}
		Following \cite{Fernandez-Alvarez-Senovilla2020b}, in the case of $\Lambda>0$ the no-radiation condition is equivalently stated as
			\begin{equation}\label{eq:condition-covariant-dS}
			\text{There is no gravitational radiation at $\scri$} \iff \ct{D}{^\mu_{\beta}}\ct{C}{_{\mu\gamma}}\ct{\eta}{^{\beta\gamma\alpha\mu}}\ct{n}{_{\mu}}=0\ .
			\end{equation}
		 where
				\begin{align}
			    \ct{C}{_{\alpha\beta}} &\defeqs\frac{1}{2} \ct{n}{^{\mu}}\ct{n}{^{\nu}}\eta_{\alpha\mu}{}^{\rho\sigma}  \ct{d}{_{\rho\sigma\beta\nu}} \ ,\label{eq:n-magnetic}\\
			    \ct{D}{_{\alpha\beta}} &\defeqs \ct{n}{^{\mu}}\ct{n}{^{\nu}}\ct{d}{_{\alpha\mu\beta\nu}}\ ,\label{eq:n-electric}
			 	\end{align}
		and $\eta_{\alpha\beta\lambda\mu}$ is the canonical volume element 4-form in $\prn{M,\ct{g}{_{\alpha\beta}}}$ and $\ct{d}{_{\alpha\beta\gamma}^\delta}$ the rescaled Weyl tensor \eqref{eq:rescaled-Weyl}. These are the standard electric and magnetic parts of the rescaled Weyl tensor with respect to the timelike unit normal $\ct{n}{^\alpha}$ \eqref{eq:unit-normal-scri} which completely determine the rescaled Weyl tensor at $\scri$. The magnetic part \eqref{eq:n-magnetic} is proportional to the Cotton-York tensor $\ct{Y}{_{\alpha\beta}} $ of $\prn{\scri,\ct{h}{_{ab}}}$ ---see, e.g., \cite{Fernandez-Senovilla:2022b}---, whereas the electric part $\ct{D}{_{\alpha\beta}}$ is part of the covariant initial data $\prn{\scri,h,D}$ at $\scri$ \cite{Friedrich1986a}. We compute them explicitly for the present metric ---see \cref{sec:C-D} for the $\Lambda<0$ case, where these tensors have also different meanings.\\
	
		Recall that close to $\scri$ one has $Q<0$ for $\Lambda>0$, and from \cref{eq:efe-scri} it also follows that  $Q+\alpha^2 P<0$. Then, define the orthonormal basis $\cbrkt{\ct{n}{^\alpha},\ct{t}{^\alpha},\ct{r}{^\alpha},\ct{s}{^\alpha}}$:
			\begin{align}
			    \ct{n}{^{\alpha}} &= \frac{1}{\rho\sqrt{-Q-\alpha^2P}}\prn{Q\delta_{q}^\alpha-\alpha P\delta^{\alpha}_{x}}\ ,\\
			    \ct{t}{^{\alpha}} &= \frac{\sqrt{C_f}}{\rho\sqrt{-Q}}\prn{D\delta^\alpha_{t}-C\delta^\alpha_{\varphi}} \ ,\\
			    \ct{r}{^{\alpha}}&=\frac{\sqrt{C_f}}{\rho\sqrt{P}}\prn{B\delta^{\alpha}_{t}+A\delta^{\alpha}_{\varphi}}\ ,\\
			    \ct{s}{^{\alpha}}&=\sqrt{\frac{PQ}{Q+\alpha^2P}}\frac{1}{\rho}\prn{\alpha\delta^{\alpha}_{q}+\delta^{\alpha}_{x}}\ .
			\end{align}
		At $\scri$, $\ct{n}{^\alpha}$ coincides with the unit normal (and it is timelike and points outwards). The triad $\cbrkt{\ct{t}{^\alpha},\ct{r}{^\alpha},\ct{s}{^\alpha}}$ forms a basis of the three-dimensional tangent space at points on $\scri$, and thus they are spacelike.
		
Using this basis and evaluating all functions at $\scri$, we get the following expressions:\\
		\subparagraph{\textbf{Charged, rotating and accelerating black hole $\Lambda>0$}\\}
		Setting $l=0$,
			\begin{align}
				    \ct{C}{_{\alpha\beta}} =& -\frac{a\alpha K}{\prn{1+a^2\alpha^2x^4}^4\prn{1+\alpha^2a^2}} \Biggl\{\brkt{\frac{9}{\Lambda}\frac{P\alpha^2}{\prn{1+\alpha^2a^2}}+2\prn{1+a^2\alpha^2x^4}}\ct{t}{_{\alpha}}\ct{t}{_{\beta}}\nonumber\\
				    &-\brkt{\frac{9}{\Lambda}\frac{P\alpha^2}{\prn{1+\alpha^2a^2}}+\prn{1+a^2\alpha^2x^4}}\ct{r}{_{\alpha}}\ct{r}{_{\beta}}-\prn{1+a^2\alpha^2x^4}\ct{s}{_{\alpha}}\ct{s}{_{\beta}}\Biggr\}\nonumber\\
				    &+\frac{\alpha L}{\prn{1+a^2\alpha^2x^4}^4}\frac{9}{\Lambda}\frac{\sqrt{-PQ}}{\prn{1+\alpha^2a^2}^2}\prn{\ct{t}{_{\alpha}}\ct{r}{_{\beta}}+\ct{t}{_{\beta}}\ct{r}{_{\alpha}}} \ ,\\
				    \ct{D}{_{\alpha\beta}} =& -\frac{L}{\prn{1+a^2\alpha^2x^4}^4\prn{1+\alpha^2a^2}} \Biggl\{\brkt{\frac{9}{\Lambda}\frac{P\alpha^2}{\prn{1+\alpha^2a^2}}+2\prn{1+a^2\alpha^2x^4}}\ct{t}{_{\alpha}}\ct{t}{_{\beta}}\nonumber\\
	   			    &-\brkt{\frac{9}{\Lambda}\frac{P\alpha^2}{\prn{1+\alpha^2a^2}}+\prn{1+a^2\alpha^2x^4}}\ct{r}{_{\alpha}}\ct{r}{_{\beta}}-\prn{1+a^2\alpha^2x^4}\ct{s}{_{\alpha}}\ct{s}{_{\beta}}\Biggr\}\nonumber\\
	   			    &-\frac{a\alpha^2K}{\prn{1+a^2\alpha^2x^4}^4}\frac{9}{\Lambda}\frac{\sqrt{-PQ}}{\prn{1+\alpha^2a^2}^2}\prn{\ct{t}{_{\alpha}}\ct{r}{_{\beta}}+\ct{t}{_{\beta}}\ct{r}{_{\alpha}}} \ ,
				\end{align}
		where
			\begin{align}
				L&\defeq  m+\alpha x\cbrkt{a^2\alpha x\brkt{\alpha^2a^2mx^4+\prn{2\alpha g^2+2\alpha e^2}x^3-3mx^2-3m}-2\prn{e^2+g^2}}\ ,\label{eq:L}\\
				K&\defeq  x^2\cbrkt{\alpha x\brkt{a^2\alpha mx^3+3a^2\alpha m x+ 4\prn{g^2+e^2}-3m}-m}\ .\label{eq:K}
			\end{align}
		\subparagraph{\textbf{Charged C-metric $\Lambda>0$}\\}
		Setting $l=a=0$,
			\begin{align}
	    \ct{C}{_{\alpha\beta}}& =\frac{9}{\Lambda}\alpha\sqrt{-PQ}\brkt{m-2\alpha\prn{e^2+g^2}x}\prn{\ct{t}{_{\alpha}}\ct{r}{_{\beta}}+\ct{t}{_{\beta}}\ct{r}{_{\alpha}}} \ ,\\
	    \ct{D}{_{\alpha\beta}} =& \brkt{m-2\alpha\prn{e^2+g^2}x} \Biggl\{-\brkt{\frac{9}{\Lambda}P\alpha^2+2}\ct{t}{_{\alpha}}\ct{t}{_{\beta}}\nonumber\\
				    &+\brkt{\frac{9}{\Lambda}P\alpha^2+1}\ct{r}{_{\alpha}}\ct{r}{_{\beta}}+1\ct{s}{_{\alpha}}\ct{s}{_{\beta}}\Biggr\}\ .
				\end{align} 
	
		\subparagraph{\textbf{C-metric $\Lambda>0$}\\}
		Setting $l=g=e=a=0$,
			\begin{align}
			    \ct{C}{_{\alpha\beta}} =&\frac{9}{\Lambda}\alpha m\sqrt{-PQ}\prn{\ct{t}{_{\alpha}}\ct{r}{_{\beta}}+\ct{t}{_{\beta}}\ct{r}{_{\alpha}}} \ ,\\
			    \ct{D}{_{\alpha\beta}} =&- m\prn{\frac{9}{\Lambda}\alpha^2P+2}\ct{t}{_{\alpha}}\ct{t}{_{\beta}}+m\prn{\frac{9}{\Lambda}\alpha^2P+1}\ct{r}{_{\alpha}}\ct{r}{_{\beta}}+m\ct{s}{_{\alpha}}\ct{s}{_{\beta}}\ .
			\end{align}
		 Notice that in this case $\delta_{t}^\alpha$ and $\delta_{\varphi}^\alpha$, the spacelike CKVFs tangent to $\scri$,  are eigenvectors of $\ct{D}{_{\alpha\beta}}$ but not of $\ct{C}{_{\alpha\beta}}$\ .
		\subparagraph{\textbf{The most general non-radiating case $\Lambda>0$}\\}
		If we set $\alpha=0$ (keeping the rest of parameters) we get non accelerating black holes with
		\begin{align}
		    \ct{C}{_{\alpha\beta}} &= l\prn{\frac{\Lambda}{3}l^2-1}\prn{3\ct{t}{_{\alpha}}\ct{t}{_{\beta}}-\ct{P}{_{\alpha\beta}}}\ , \\
		    \ct{D}{_{\alpha\beta}} &=- m\prn{3\ct{t}{_{\alpha}}\ct{t}{_{\beta}}-\ct{P}{_{\alpha\beta}}}\ .
		\end{align}
		where the projector to $\scri$, $\ct{P}{_{\alpha\beta}}=\ct{t}{_{\alpha}}\ct{t}{_{\beta}}+\ct{r}{_{\alpha}}\ct{r}{_{\beta}}+\ct{s}{_{\alpha}}\ct{s}{_{\beta}}$ has been introduced. Therefore,
			\begin{equation}
			l\prn{1-\frac{\Lambda}{3}l^2}\ct{D}{_{\alpha\beta}}= 	m\ct{C}{_{\alpha\beta}}\ .
			\end{equation}
		 It is obvious that these expressions satisfy the non-radiation condition. In fact, this is a non-conformally-flat case, i.e., $\ct{C}{_{\alpha\beta}}$ is different from zero. This illustrates the results of \cite{Fernandez-Alvarez-Senovilla2020b,Fernandez-Senovilla:2022b}, and departs from the typical constraint of imposing conformal flatness to remove gravitational radiation; there are other non-radiating scenarios possible. There is a special combination though,
			\begin{equation}\label{eq:conf-flat-NUT}
			\frac{\Lambda}{3}l^2=1\ ,
			\end{equation}
		which is a conformally flat subcase. One further observation is in order. Observe that with $\alpha=0$ $\ct{t}{^\alpha}$ is  proportional to the spacelike CKVF $\ct{\xi_t}{^{\alpha}}=\partial^\alpha_{t}$ at $\scri$,
		\begin{equation}
		\sqrt{\frac{-Q}{C_f}}\ct{t}{^\alpha}=\sqrt{\frac{\Lambda}{3 C_f}}\ct{t}{^{\alpha}}=\ct{\xi_t}{^{\alpha}},	
	\end{equation}
		and $ Q=-\Lambda/3 $ at $ \scri $ for $ \alpha=0 $.
		 This CKVF is an eigenvector of both $\ct{C}{_{\alpha\beta}}$ and $\ct{D}{_{\alpha\beta}}$. Using three-dimensional indices $a, b, c...$ within $\scri$, we can write the pullback of the expressions above to get
			\begin{align}
			    \ct{C}{_{ab}} =& \frac{l\prn{\Lambda l^2-3}}{\abs{\xi_t}^2} \prn{\ct{\xi_t}{_{a}}\ct{\xi_t}{_{b}}-\frac{\abs{\xi_t}^2}{3}\ct{h}{_{ab}}}\ ,\label{eq:C-norad-Qpos}\\
			    \ct{D}{_{ab}} =&  -\frac{3m}{\abs{\xi_t}^2} \prn{\ct{\xi_t}{_{a}}\ct{\xi_t}{_{b}}-\frac{\abs{\xi_t}^2}{3}\ct{h}{_{ab}}}\ .\label{eq:D-norad-Qpos}
			\end{align}
		We have introduced ${\xi_{t}}^2\defeq{\ct{\xi_{t}}{_{\mu}}\ct{\xi_{t}}{^{\mu}}}=\ct{\xi_{t}}{_{\mu}}\ct{\xi_{t}}{^{\mu}}=\Lambda/3 C_f$. This agrees with results in \cite{MarsSenovilla:2013,MarsPaetzSenovilla2017}, and the conformally flat subcase given by \cref{eq:conf-flat-NUT} has the same  structure found in the classification of Kerr-de Sitter-like spacetimes with conformally flat $\scri$ in \cite{MarsPaetzSenovilla2017} ---up to a different choice of gauge. We will see that the analogous structure appears when $\Lambda<0$ ---see \cref{sec:C-D}. 

\section{Gravitational radiation with  ${\Lambda<0}$}\label{sec:radiation-Lambda-n}

First we characterize the asymptotic gravitational radiation arriving at (or departing from) $\scri$  with $\Lambda<0$. To this end, we use the characterization given in \cite{FernandezSenovilla:2026} which can be stated in terms of asymptotic super-Poynting vector fields or equivalently in terms of algebraic structure of the rescaled Weyl tensor. We will start with the $\bar{Q} >0$ regions.\\

 Any unit timelike vector field $u^\alpha$ tangent to $\scri$ can be expressed using real positive functions $a$, $b$ and a complex one $d$ as
	\begin{equation}
	u^\alpha = ak^\alpha  +bl^\alpha + dm^\alpha + \bar{d}\bar{m}^\alpha\ ,
	\label{general-oberver-u}
	\end{equation}
with the constraints
	\begin{equation}\label{eq:constraints}
	    ab-d\bar{d} = \frac{1}{2}\ ,\quad aB_{1}+bB_2+dE+d\bar{E}=0\ .
	\end{equation}
Here $B_1$, $B_2$, $E$ are
	\begin{equation}\label{eq:contractions}
	    B_1   \defeq k^\mu n_{\mu}\  ,\quad  B_2 \defeq l^\mu n_\mu\ , \quad E \defeq  m^\mu n_{\mu}\ .
	\end{equation}
Using \cref{eq:unit-normal-scri}, the explicit form of these contractions is 
\begin{align}	\label{eq:PND-orientations-Q>0}
 B_1 &= -B_2 = -\sqrt{\frac{3}{-\Lambda}}\frac{\sqrt{\bar{Q}}}{\sqrt 2\,\bar{\rho}}\ ,\\
 E&= i \alpha \sqrt{\frac{3}{-\Lambda}}\frac{\sqrt{P}}{\sqrt 2\,\bar{\rho}}\ ,
\end{align}
Observe that since $B_1=-B_2$, the two PNDs are oriented oppositely at $\scri$:  $k^\alpha$ points outwards, while  $l^\alpha$ points inwards.\\

Theorem 2 of \cite{FernandezSenovilla:2026} states for type-D spacetimes that absence of radiation at $\scri$ is equivalent to either the coplanarity of the normal $\ct{n}{^\alpha}$ and the 2 PNDs, or to both PNDs tangent to $\scri$ ( $B_1=0=B_2$ ). Inspecting \cref{eq:PND-orientations-Q>0}, it is clear that the latter is impossible (except at points where $\bar{Q}$ has zeros), whereas from \cref{eq:normal-scri} and \eqref{nullframe-Q>0} we deduce that coplanarity holds if and only if $\alpha=0$. In other words, we have shown that
\begin{center}
 \emph{there is no gravitational radiation if and only if the acceleration parameter $\alpha$ vanishes}.
\end{center} Actually, Theorem 2 is an implication of the general criterion
	\begin{equation*}
	\text{There is no gravitational radiation }\iff \ct{\overline{\P}}{^\alpha}\prn{u}\ct{n}{_{\alpha}} = 0 \quad \forall \ct{u}{^\alpha}\ .
	\end{equation*}
Here $\ct{\overline{\P}}{^\alpha}\prn{u}$ is the asymptotic s-Poynting vector field of observer $\ct{u}{^\alpha}$.  We can  directly employ formula (B.58) of \cite{FernandezSenovilla:2026}, simplified for type D spacetimes:
\begin{equation}\label{eq:s-Poynting-n}
\ct{\overline{\P}}{^\mu}({u})\ct{n}{_{\mu}} = 36 \,\phi_2 \bar{\phi}_2\,(ab+d\bar{d}\,)\big( a\,B_1 + b\,B_2 \big),
\end{equation}
and using \cref{eq:PND-orientations-Q>0},
\begin{equation}\label{eq:sP-Qpos}
\ct{\overline{\P}}{^\mu}({u})\ct{n}{_{\mu}} = -36\sqrt{\frac{3}{-\Lambda}}\frac{\sqrt{\bar{Q}}}{\sqrt 2\,\bar{\rho}} \,|\phi_2|^2\,(ab+d\bar{d}\,)\big( a - b \big)\ .
\end{equation}
	If $\alpha\neq 0$, there exist some $\ct{u}{^\alpha}$ satisfying $a-b\neq 0$ and constraint \eqref{eq:constraints}, therefore, \emph{gravitational radiation is present}. If, on the contrary, $\alpha=0$, automatically this sets $D=0$, and by the second of \cref{eq:constraints} and \cref{eq:PND-orientations-Q>0} it follows that $a-b=0$ for all $\ct{u}{^\alpha}$ tangent to $\scri$: \emph{no gravitational radiation}. Hence, we have reached the same conclusion. Additionally, notice that the sign of \cref{eq:sP-Qpos} for each observer  depends on the coefficients $a$ and $b$: an observer with $a>b$ (i.e., tilted in the $\ct{k}{^\alpha}$-direction) will experience a transverse flux of superenergy that arrives at $\scri$ from the space-time (which agrees with the orientation of $\ct{{k}}{^\alpha}$).\\

If one wishes to consider a region on $\scri$ with $\bar{Q} < 0$ so that necessarily $\alpha\neq 0$ and also, according to the discussion in previous sections, we are on a region without roots of $P(x)$, on  using \cref{nullframe-Q<0} one has
\begin{align}	\label{eq:PND-orientations-Q<0}
 B_1 &= B_2 = -\sqrt{\frac{3}{-\Lambda}}\frac{\sqrt{-\bar{Q}}}{\sqrt 2\,\bar{\rho}}\ ,\\
 E&= i \alpha \sqrt{\frac{3}{-\Lambda}}\frac{\sqrt{\bar{P}}}{\sqrt 2\,\bar{\rho}}\ ,
\end{align}
showing that both principal null directions  \emph{point outwards}. The orthogonal component of the asymptotic super-Poynting for an arbitrary timelike observer $\ct{u}{^{\alpha}}$ tangent to $\scri$ \eqref{general-oberver-u} now reads
\begin{equation}
\ct{\overline{\P}}{^\mu}({u})\ct{n}{_{\mu}} =  -36\sqrt{\frac{3}{-\Lambda}}\frac{\sqrt{-\bar{Q}}}{\sqrt 2\,\bar{\rho}} \,\phi_2\bar{\phi}_2\,\prn{ab+d\bar{d}}\big( a + b \big)<0\ .
\end{equation}
The sign shows that the transversal flux of super-energy opposes $\ct{n}{^{\alpha}}$, i.e., it escapes from space-time, which agrees with the orientation of the two PNDs.

	\subsection{The Cotton-York and the holographic stress-energy tensors}\label{sec:C-D}
	As proved in Theorem 1 of \cite{FernandezSenovilla:2026}, the no-radiation condition with $\Lambda < 0$ is equivalently stated as
		\begin{equation}\label{eq:condition-covariant}
		\text{There is no gravitational radiation at $\scri$} \iff \beta\ct{D}{_{\alpha\beta}}=\gamma\ct{C}{_{\alpha\beta}}
		\end{equation}
		with $\beta$ and $\gamma$  functions such that $\beta^2+\gamma^2\neq 0$. These tensors are defined as in \cref{eq:n-electric,eq:n-magnetic}, but note that in this case the normal $\ct{n}{^{\alpha}}$ is spacelike and points inwards. As in the case $\Lambda>0$ ---see \cref{sec:C-D-dS}--- the tensor $\ct{C}{_{\alpha\beta}}$ is proportional to the Cotton-York tensor of $\prn{\scri,\ct{h}{_{ab}}}$, and  $\ct{D}{_{\alpha\beta}}$ is often called the holographic stress-energy tensor of the boundary in the context of holography \cite{Balasubramanian1999}. At the same time, they are related to the initial boundary data problem \cite{Friedrich1995} ---see \cite{FernandezSenovilla:2026} and references therein for further discussion of these aspects.  
	\subsubsection{$Q>0$ case}
	Let us define the following orthonormal basis $\cbrkt{\ct{t}{^\alpha},\ct{n}{^\alpha},\ct{r}{^\alpha},\ct{s}{^\alpha}}$:
		\begin{align}
		    \ct{t}{^{\alpha}} &= \frac{\sqrt{C_f}}{\rho\sqrt{Q}}\prn{D\delta^\alpha_{t}-C\delta^\alpha_{\varphi}} \ ,\\
		    \ct{n}{^{\alpha}} &= \frac{1}{\rho\sqrt{Q+\alpha^2P}}\prn{Q\delta_{q}^\alpha-\alpha P\delta^{\alpha}_{x}}\ ,\\
		    \ct{r}{^{\alpha}}&=\frac{\sqrt{C_f}}{\rho\sqrt{P}}\prn{B\delta^{\alpha}_{t}+A\delta^{\alpha}_{\varphi}}\ ,\\
		    \ct{s}{^{\alpha}}&=\sqrt{\frac{PQ}{Q+\alpha^2P}}\frac{1}{\rho}\prn{\alpha\delta^{\alpha}_{q}+\delta^{\alpha}_{x}}\ .
		\end{align}
	At $\scri$, $\ct{n}{^\alpha}$ is just the unit normal. Thus, $\cbrkt{\ct{t}{^\alpha},\ct{r}{^\alpha},\ct{s}{^\alpha}}$ forms a basis on $\scri$, with $\ct{t}{^\alpha}$ being timelike and the other two, spacelike. A direct calculation gives  $\ct{C}{_{\alpha\beta}}$ and $\ct{D}{_{\alpha\beta}}$ ---all functions are evaluated at $\scri$:\\
	\subparagraph{\textbf{Charged, rotating and accelerating black hole $\Lambda<0$, $Q>0$}\\}
	Setting $l=0$,
		\begin{align}
			    \ct{C}{_{\alpha\beta}} =& -\frac{a\alpha K}{\prn{1+a^2\alpha^2x^4}^4\prn{1+\alpha^2a^2}} \Biggl\{\brkt{\frac{9}{\Lambda}\frac{P\alpha^2}{\prn{1+\alpha^2a^2}}+2\prn{1+a^2\alpha^2x^4}}\ct{t}{_{\alpha}}\ct{t}{_{\beta}}\nonumber\\
			    &+\brkt{\frac{9}{\Lambda}\frac{P\alpha^2}{\prn{1+\alpha^2a^2}}+\prn{1+a^2\alpha^2x^4}}\ct{r}{_{\alpha}}\ct{r}{_{\beta}}+\prn{1+a^2\alpha^2x^4}\ct{s}{_{\alpha}}\ct{s}{_{\beta}}\Biggr\}\nonumber\\
			    &-\frac{\alpha L}{\prn{1+a^2\alpha^2x^4}^4}\frac{9}{\Lambda}\frac{\sqrt{PQ}}{\prn{1+\alpha^2a^2}^2}\prn{\ct{t}{_{\alpha}}\ct{r}{_{\beta}}+\ct{t}{_{\beta}}\ct{r}{_{\alpha}}} \ ,\\
			    \ct{D}{_{\alpha\beta}} =& -\frac{L}{\prn{1+a^2\alpha^2x^4}^4\prn{1+\alpha^2a^2}} \Biggl\{\brkt{\frac{9}{\Lambda}\frac{P\alpha^2}{\prn{1+\alpha^2a^2}}+2\prn{1+a^2\alpha^2x^4}}\ct{t}{_{\alpha}}\ct{t}{_{\beta}}\nonumber\\
   			    &+\brkt{\frac{9}{\Lambda}\frac{P\alpha^2}{\prn{1+\alpha^2a^2}}+\prn{1+a^2\alpha^2x^4}}\ct{r}{_{\alpha}}\ct{r}{_{\beta}}+\prn{1+a^2\alpha^2x^4}\ct{s}{_{\alpha}}\ct{s}{_{\beta}}\Biggr\}\nonumber\\
   			    &+\frac{a\alpha^2K}{\prn{1+a^2\alpha^2x^4}^4}\frac{9}{\Lambda}\frac{\sqrt{PQ}}{\prn{1+\alpha^2a^2}^2}\prn{\ct{t}{_{\alpha}}\ct{r}{_{\beta}}+\ct{t}{_{\beta}}\ct{r}{_{\alpha}}} \ ,
			\end{align}
		where $L$ and $K$ are defined in \cref{eq:L,eq:K}. The no-radiation condition $\eqref{eq:condition-covariant}$ is violated unless $K=0$ and $L=0$, which may hold for some values of $x$ but it is the trivial case $\ct{d}{_{\alpha\beta\gamma}^{\delta}}=0$. Thus, the only possibility is $\alpha=0$.
	\subparagraph{\textbf{Charged C-metric $\Lambda<0$, $Q>0$}\\}
	Setting $l=a=0$,
		\begin{align}
    \ct{C}{_{\alpha\beta}}& =-\frac{9}{\Lambda}\alpha\sqrt{PQ}\brkt{m-2\alpha\prn{e^2+g^2}x}\prn{\ct{t}{_{\alpha}}\ct{r}{_{\beta}}+\ct{t}{_{\beta}}\ct{r}{_{\alpha}}} \ ,\\
    \ct{D}{_{\alpha\beta}} =& -\brkt{m-2\alpha\prn{e^2+g^2}x} \Biggl\{\brkt{\frac{9}{\Lambda}P\alpha^2+2}\ct{t}{_{\alpha}}\ct{t}{_{\beta}}\nonumber\\
			    &+\brkt{\frac{9}{\Lambda}P\alpha^2+1}\ct{r}{_{\alpha}}\ct{r}{_{\beta}}+\ct{s}{_{\alpha}}\ct{s}{_{\beta}}\Biggr\}\ .
			\end{align} 
	As in the previous case, the no-radiation condition requires $m-2\alpha\prn{e^2+g^2}x=0$ but this corresponds to the Petrov type 0 case. Then, one falls within the $\alpha=0$ case again.
	\subparagraph{\textbf{C-metric $\Lambda<0$, $Q>0$}\\}
	Setting $l=g=e=a=0$,
		\begin{align}
		    \ct{C}{_{\alpha\beta}} =& -\frac{9}{\Lambda}\alpha m\sqrt{P}\sqrt{Q}\prn{\ct{t}{_{\alpha}}\ct{r}{_{\beta}}+\ct{t}{_{\beta}}\ct{r}{_{\alpha}}} \ ,\\
		    \ct{D}{_{\alpha\beta}} =&- m\prn{\frac{9}{\Lambda}\alpha^2P+2}\ct{t}{_{\alpha}}\ct{t}{_{\beta}}-m\prn{\frac{9}{\Lambda}\alpha^2P+1}\ct{r}{_{\alpha}}\ct{r}{_{\beta}}-m\ct{s}{_{\alpha}}\ct{s}{_{\beta}}\ .
		\end{align}
	 It is easy to see that in this case condition \eqref{eq:condition-covariant} happens if and only if $\alpha=0$, in which case, $\beta=0$ and $\ct{C}{_{\alpha\beta}}=0$, thus $\ct{h}{_{ab}}$ is conformally flat. Hence, for the C-metric there is always radiation. Observe that $\partial_{t}^\alpha$, the timelike CKVF tangent to $\scri$,  is an eigenvector of $\ct{D}{_{\alpha\beta}}$ but not of $\ct{C}{_{\alpha\beta}}$\ .
	\subparagraph{\textbf{The most general non-radiating case $\Lambda<0$, $Q>0$}\\}
	If we set $\alpha=0$ (keeping the rest of parameters) we get non accelerating black holes with
	\begin{align}
	    \ct{C}{_{\alpha\beta}} &= l\prn{\frac{\Lambda}{3}l^2-1}\prn{3\ct{t}{_{\alpha}}\ct{t}{_{\beta}}+\ct{P}{_{\alpha\beta}}}\ , \\
	    \ct{D}{_{\alpha\beta}} &=- m\prn{3\ct{t}{_{\alpha}}\ct{t}{_{\beta}}+\ct{P}{_{\alpha\beta}}}\ .
	\end{align}
	Here $\ct{P}{_{\alpha\beta}}=-\ct{t}{_{\alpha}}\ct{t}{_{\beta}}+\ct{r}{_{\alpha}}\ct{r}{_{\beta}}+\ct{s}{_{\alpha}}\ct{s}{_{\beta}}$ is the projector to $\scri$, i.e., both tensors are diagonal in this basis. Therefore,
		\begin{equation}
		l\prn{1-\frac{\Lambda}{3}l^2}\ct{D}{_{\alpha\beta}}= 	m\ct{C}{_{\alpha\beta}}\ ,
		\end{equation}
	meaning that there is no radiation. Observe that this time one has $\beta=	l\prn{\frac{\Lambda}{3}l^2-1}$ and $\gamma=-m$ and that they are constant. Condition \eqref{eq:condition-covariant}, however, allows for $\beta$ and $\gamma$ functions, but such a case does not appear in the present family of metrics.
	Also, $\ct{C}{_{\alpha\beta}}$ is different from zero, so these are spacetimes with a \textbf{non-}conformally-flat metric at $\scri$. This nicely agrees with theorem 1 in \cite{FernandezSenovilla:2026}. Observe that the sign of the cosmological constant implies that
		\begin{equation}
			\beta=0 \iff l=0
		\end{equation}
	 Hence, we arrive at the conclusion that, within this family of metrics, \emph{there are no conformally flat cases unless $l=0$}. This departs from the scenario with $\Lambda>0$ ---see \cref{sec:C-D-dS}.
	 
	Similarly as before, if $\alpha=0$ $\ct{t}{^\alpha}$ is  proportional to the timelike CKVF $\ct{\xi_t}{^{\alpha}}=\partial^\alpha_{t}$ at $\scri$,
	\begin{equation}
	\sqrt{\frac{-\Lambda}{3 C_f}}\ct{t}{^\alpha}=\ct{\xi_t}{^{\alpha}}.
	\end{equation}
	 Again, this CKVF is an eigenvector of  $\ct{C}{_{\alpha\beta}}$ and $\ct{D}{_{\alpha\beta}}$ and one can write
		\begin{align}
		    \ct{C}{_{ab}} =& \frac{3}{\abs{\xi_t}^2} \beta\prn{\ct{\xi_t}{_{a}}\ct{\xi_t}{_{b}}+\frac{\abs{\xi_t}^2}{3}\ct{h}{_{ab}}}\ ,\label{eq:C-norad-Qpos-ads}\\
		    \ct{D}{_{ab}} =&  \frac{3}{\abs{\xi_t}^2} \gamma\prn{\ct{\xi_t}{_{a}}\ct{\xi_t}{_{b}}+\frac{\abs{\xi_t}^2}{3}\ct{h}{_{ab}}}\ .\label{eq:D-norad-Qpos-ads}
		\end{align}
	This time we define $\abs{\xi_{t}}^2\defeq\abs{\ct{\xi_{t}}{_{\mu}}\ct{\xi_{t}}{^{\mu}}}=-\ct{\xi_{t}}{_{\mu}}\ct{\xi_{t}}{^{\mu}}=-\Lambda/3C_f$. 	As a further observation, \cref{eq:C-norad-Qpos-ads,eq:D-norad-Qpos-ads} take the form of a perfect-Cotton geometry and a holographic stress tensor in perfect equilibrium, respectively, as defined and studied in \cite{MukhopadhyayEtal2014}.
	
	\subsubsection{$Q<0$ case}
	This is the case with $\alpha\neq0$ always. We define our orthonormal basis $\cbrkt{\ct{s}{^\alpha},\ct{n}{^\alpha},\ct{r}{^\alpha},\ct{t}{^\alpha}}$ as:
		\begin{align}
		    \ct{t}{^{\alpha}} &= \frac{\sqrt{C_f}}{\rho\sqrt{-Q}}\prn{D\delta^\alpha_{t}-C\delta^\alpha_{\varphi}} \ ,\\
		    \ct{n}{^{\alpha}} &= \frac{1}{\rho\sqrt{Q+\alpha^2P}}\prn{Q\delta_{q}^\alpha-\alpha P\delta^{\alpha}_{x}}\ ,\\
		    \ct{r}{^{\alpha}}&=\frac{\sqrt{C_f}}{\rho\sqrt{P}}\prn{B\delta^{\alpha}_{t}+A\delta^{\alpha}_{\varphi}}\ ,\\
		    \ct{s}{^{\alpha}}&=\sqrt{\frac{-PQ}{Q+\alpha^2P}}\frac{1}{\rho}\prn{\alpha\delta^{\alpha}_{q}+\delta^{\alpha}_{x}}\ .
		\end{align}	
	Notice that here $\ct{t}{^\mu}\ct{t}{_{\mu}}=1$ whereas $\ct{s}{^{\mu}}\ct{s}{_{\mu}}=-1$, and recall that the KVF $\partial^\alpha_{t}$ now is spacelike ---see \cref{eq:normKVFs-2}.
		\subparagraph{\textbf{Charged, rotating and accelerating black hole $\Lambda<0$, $Q<0$}\\}
		Setting $l=0$,
			\begin{align}
				    \ct{C}{_{\alpha\beta}} =& \frac{a\alpha K}{\prn{1+a^2\alpha^2}\prn{1+a^2\alpha^2x^4}^4} \Biggl\{\brkt{\frac{9}{\Lambda}\frac{P\alpha^2}{\prn{1+a^2\alpha^2}}+2\prn{1+a^2\alpha^2x^4}}\ct{t}{_{\alpha}}\ct{t}{_{\beta}}\nonumber\\
				    &-\brkt{\frac{9}{\Lambda}\frac{P\alpha^2}{\prn{1+a^2\alpha^2}}+\prn{1+a^2\alpha^2x^4}}\ct{r}{_{\alpha}}\ct{r}{_{\beta}}+\prn{1+a^2\alpha^2x^4}\ct{s}{_{\alpha}}\ct{s}{_{\beta}}\Biggr\}\nonumber\\
				    &-\frac{\alpha L}{\prn{1+a^2\alpha^2x^4}^4}\frac{9}{\Lambda}\frac{\sqrt{-PQ}}{\prn{1+a^2\alpha^2}^2}\prn{\ct{t}{_{\alpha}}\ct{r}{_{\beta}}+\ct{t}{_{\beta}}\ct{r}{_{\alpha}}} \ ,\\
				    \ct{D}{_{\alpha\beta}} =& \frac{L}{\prn{1+a^2\alpha^2}\prn{1+a^2\alpha^2x^4}^4} \Biggl\{\brkt{\frac{9}{\Lambda}\frac{P\alpha^2}{\prn{1+a^2\alpha^2}}+2\prn{1+a^2\alpha^2x^4}}\ct{t}{_{\alpha}}\ct{t}{_{\beta}}\nonumber\\
	   			    &-\brkt{\frac{9}{\Lambda}\frac{P\alpha^2}{\prn{1+a^2\alpha^2}}+\prn{1+a^2\alpha^2x^4}}\ct{r}{_{\alpha}}\ct{r}{_{\beta}}+\prn{1+a^2\alpha^2x^4}\ct{s}{_{\alpha}}\ct{s}{_{\beta}}\Biggr\}\nonumber\\
	   			    &+\frac{a\alpha^2K}{\prn{1+a^2\alpha^2x^4}^4}\frac{9}{\Lambda}\frac{\sqrt{-PQ}}{\prn{1+a^2\alpha^2}^2}\prn{\ct{t}{_{\alpha}}\ct{r}{_{\beta}}+\ct{t}{_{\beta}}\ct{r}{_{\alpha}}} \ ,
				\end{align}
		where $L$ and $K$ are defined in \cref{eq:L,eq:K}. As before, the no-radiation condition only holds for the case $K=0$ and $L=0$, i.e., $\ct{d}{_{\alpha\beta\gamma}^{\delta}}=0$.
		\subparagraph{\textbf{Charged C-metric $\Lambda<0$, $Q<0$}\\}
		Setting $l=a=0$,
			\begin{align}
	    \ct{C}{_{\alpha\beta}}& =\frac{9}{\Lambda}\alpha\sqrt{-PQ}\brkt{m-2\alpha\prn{e^2+g^2}x}\prn{\ct{t}{_{\alpha}}\ct{r}{_{\beta}}+\ct{t}{_{\beta}}\ct{r}{_{\alpha}}} \ ,\\
	    \ct{D}{_{\alpha\beta}} =& \brkt{m-2\alpha\prn{e^2+g^2}x} \Biggl\{\brkt{\frac{9}{\Lambda}P\alpha^2+2}\ct{t}{_{\alpha}}\ct{t}{_{\beta}}\nonumber\\
				    &-\brkt{\frac{9}{\Lambda}P\alpha^2+1}\ct{r}{_{\alpha}}\ct{r}{_{\beta}}+\ct{s}{_{\alpha}}\ct{s}{_{\beta}}\Biggr\}\ .
				\end{align} 
	The no-radiation condition imposes $m-2\alpha\prn{e^2+g^2}x=0$, i.e., the Petrov type 0 case.
		\subparagraph{\textbf{C-metric $\Lambda<0$, $Q<0$}\\}
		Setting $l=g=e=a=0$,
			\begin{align}
			    \ct{C}{_{\alpha\beta}} =& \frac{9}{\Lambda}\alpha m\sqrt{-PQ}\prn{\ct{t}{_{\alpha}}\ct{r}{_{\beta}}+\ct{t}{_{\beta}}\ct{r}{_{\alpha}}} \ ,\\
			    \ct{D}{_{\alpha\beta}} =& m\prn{\frac{9}{\Lambda}\alpha^2P+2}\ct{t}{_{\alpha}}\ct{t}{_{\beta}}-m\prn{\frac{9}{\Lambda}\alpha^2P+1}\ct{r}{_{\alpha}}\ct{r}{_{\beta}}+m\ct{s}{_{\alpha}}\ct{s}{_{\beta}}\ .
			\end{align}	
As a final observation, at $x=0$, $\ct{C}{_{\alpha\beta}}$ and $\ct{D}{_{\alpha\beta}}$ of the charged, rotating and accelerating black hole coincide with those at $x=0$ of the accelerating C-metric.

\section{Conclusions}
We summarize our results for the general Pleba\'nski-Demia\'nski type~D solution with cosmological constant $\Lambda$ of any sign:
	\begin{itemize}
		\item There is gravitational radiation at infinity if and only if the acceleration parameter $ \alpha $ does not vanish --- \cref{sec:radiation-Lambda-p,sec:radiation-Lambda-n}.
		\item By doing an analysis of the roots of the quartic $ P(x) $, we determine the possible intervals allowed for the range of coordinate $ x $ ---\cref{subsec: P>0}.
		\item The conditions for the existence of symmetry axes are computed, showing that ---within the allowed ranges of $ x $--- conical singularities may or may not be present. We also determine how the axis can be regularised and in which cases. Then,  for the case of a unique axial KVF, one of the components of the axis has a non-zero deficit angle if and only if there is gravitational radiation at infinity---\cref{subsec:cones}. 
		\item The Cotton-York tensor and the boundary stress-energy tensor and initial and final data tensor $ D $ are computed for several cases. For the most general non-radiating case ($ \alpha=0 $) we show the conditions for having a conformally flat metric at $ \scri $. For $ \Lambda<0 $ this is only possible if $ l=0 $ ---\cref{sec:C-D-dS,sec:C-D}.
		\item {\blue} The general non-radiating solutions with non-vanishing Cotton-York tensor have been identified for $ \Lambda>0 $ and $ \Lambda<0 $.
	\end{itemize}
	
	\subsection*{Acknowledgments}
FFA was partly supported by a contract of the project PID2021-123226NB-I00, financed by the Spanish MICIU/AEI and FEDER (European Union). The authors thank J. Podolský for useful discussions. 

\appendix

\section{Identification of $\Lambda$ and the conical constant $C_f$}\label{sec:metric-correction}

	In the form in which the metric  was presented in \cite{OvcharenkoPodolskyAstorino:2024}, the cosmological constant was taken as $\Lambda/\prn{1+a^2\alpha^2}$, where this $\Lambda$ is the one used in the present work. That choice only provides a solution if $C_f$ is fixed to  $C_f=1/\prn{1+a^2\alpha^2}$. However, the general metric \cite{Astorino:2024b} allows for arbitrary (positive) constant $C_f$ if the cosmological constant is identified correctly. Paradoxically, the $C_f$ was treated as a free normalisation constant in \cite{Astorino:2024b,OvcharenkoPodolskyAstorino:2024}, where it is said that for convenience it can be set to the value just mentioned, even though it was assumed to be related to conicity somehow too. Moreover, in \cite{OvcharenkoPodolskyAstorino2025} it is said that $C_f$ has to be fixed to warrant the appropriate asymptotic behaviour of the metric. These statements are therefore misleading. We show in this paper that the constant $C_f$ cannot be absorbed by a homothety nor fixed beforehand, and we clarify its role in the conicity of axes in \cref{subsec:cones}. If one identifies the proper $\Lambda$, the EFE are satisfied and the conformal Einstein Field Equations CEFE too, ensuring that everything has the correct behaviour asymptotically, and $C_f$ is kept unfixed, as any other parameter of the metric. All of this  follows from the following analysis. \\

Taking the metric of eq.(25) in \cite{OvcharenkoPodolskyAstorino:2024} as starting point, the first step is to correctly identify $\Lambda$. Calling $\Lambda'$ and $g'$ the cosmological constant and metric of \cite{OvcharenkoPodolskyAstorino:2024}, respectively, one has to do the change:
	\begin{equation}
		\Lambda' \longrightarrow \Lambda''= \frac{\Lambda'}{C_f\prn{1+a^2\alpha^2}}
	\end{equation}
Then, the result of correcting $g'$ with $\Lambda''$, is a solution  of the EFE, without fixing $C_f$.  For our purposes, we further perform a homothety of the metric $g'$ to a new metric $g$
		\begin{equation}\label{eq:homothety}
			g=\frac{1}{C_f} g'\ .
		\end{equation}
Of course, after dilating the metric one has to do the corresponding rescaling in $\Lambda''$ to ensure that the metric is mapped to a solution of EFEs:
	\begin{equation}
		\Lambda'' \longrightarrow \Lambda={C_f}\Lambda'' \ .
	\end{equation}
This new metric $g$ is  presented in \cref{A-new-metric}. It is a solution of EFEs with metric functions \eqref{Omega_new}--\eqref{eq:P-hom}, $\Lambda$ and $C_f$. The whole point of the homothety is to decouple $C_f$ from the quartics $Q$ and $P$ that in their final version read as in \cref{eq:Q-hom,eq:P-hom}. To correctly map the solution, one also has to rescale accordingly the vector potential\footnote{In addition, there is a gauge function $\omega_c$ in \cite{OvcharenkoPodolskyAstorino:2024} which should have been fixed to $0$ accordingly, not to $\omega_c=-a$ as it is suggested in that work. Otherwise, the metric and the vector potential are written in different gauges.}. This is also presented in \cref{eq:vector-potential-t,eq:vector-potential-phi}.
\printbibliography
\end{document}